\documentclass[aps,reprint,pre]{revtex4-1}
\usepackage{graphicx}
\usepackage{subfigure}
\usepackage[table,dvipsnames]{xcolor}
\usepackage{longtable}
\usepackage{multirow}
\usepackage{epstopdf, epsfig}
\usepackage{amsmath,esint}
\usepackage{amsfonts}
\usepackage{dsfont}
\usepackage{dcolumn}
\usepackage{bm}
\usepackage{wasysym}
\usepackage[colorlinks=true, linkcolor=blue,citecolor=blue]{hyperref}

\DeclareSymbolFont{pxlettersA}{U}{pxmia}{m}{it}
\SetSymbolFont{pxlettersA}{bold}{U}{pxmia}{bx}{it}

\DeclareMathSymbol{\varg}{\mathord}{pxlettersA}{49}

\DeclareMathAlphabet{\mathpzc}{OT1}{pzc}{m}{it}

\setcitestyle{square}

\begin{document}

\title{Energy and momentum conservation upon reflection of a solitary pulse in a bounded magnetized plasma}
\author{Renaud Gueroult}
\affiliation{LAPLACE, Universit\'{e} de Toulouse, CNRS, INPT, UPS, 31062 Toulouse, France}

\begin{abstract}
When the nature of a magnetosonic pulse propagating in a bounded magnetized plasma slab is successively transformed from compression to rarefaction and vice-versa upon reflection at a plasma-vacuum interface, both the energy and the longitudinal electromagnetic (EM) momentum of the plasma-pulse system are found to oscillate between two states. Simple analytical models and particle-in-cell simulations show that these oscillations are associated with EM radiation to and from the surrounding magnetized vacuum.  For partial reflection supplemental losses in total pulse energy and mechanical momentum are identified and shown to follow respectively Fresnel's transmission coefficients for the energy and the magnetic perturbation. This mechanical momentum loss upon partial reflection is traced to the momentarily non-zero volume integrated Lorentz force, which in turn supports that mechanical and EM momentum transfers are respectively associated with the magnetic and electric parts of the momentum flux density.
\end{abstract}

\date{\today}

\maketitle

\section{Introduction}

The Korteweg-de-Vries (KdV) equation~\cite{Korteweg1895} describes the propagation of certain non-linear waves in weakly dispersive media for which dispersion balances the wave-steepening effects that arise from non-linearity. One particular type of solutions of the KdV equation are solitary localized waves referred to as solitons~\cite{Zabusky1965}. In unmagnetized plasmas, KdV equation have been derived for ion-acoustic (IA) waves~\cite{Washimi1966}, and the existence of ion-acoustic solitons was soon after verified in laboratory experiments~\cite{Ikezi1970}. In magnetized plasmas, KdV equations have been similarly derived for magnetosonic (MS) waves~\cite{Gardner1965,Toida1994}, and solitary waves matching the property of slow-mode MS solitons were later observed in space~\cite{Stasiewicz2003}. Beyond their basic interest, an important motivation for the study of nonlinear nonlinear MS waves (see, \emph{e.~g.}, Refs.~\cite{Mikhailovskii1985,Ohsawa2014}) has been the realisation that MS solitons can describe the initial state of the formation of subcritical perpendicular shocks~\cite{Sagdeev1966,Tidman1971,Gueroult2017}.

In a finite extent plasma, solitons can get reflected at interfaces. It has for instance been shown in an unmagnetized plasma that the sheath formed in front of a biased electrode can lead to a strong reflection of an incident IA soliton~\cite{Dahiya1978,Nishida1984,Imen1987}. Building on this finding, it was then demonstrated that an IA soliton can be forced to bounce back and forth by applying suitable boundary conditions in a laboratory plasma~\cite{Oertl1988,Cooney1991}. A similar bouncing dynamics has more recently been observed by the the author and co-workers in particle-in-cell simulations of the propagation of a MS soliton in a magnetized plasma slab bounded by vacuum~\cite{Gueroult2018b}. 

A fundamental question in these dynamical systems is energy and momentum conservation. Nagasawa and Nishida observed that the geometry of reflection of IA solitons at oblique incidence appears to fit Snell's law constructed from the transmitted and incident soliton speeds~\cite{Nagasawa1986}. Considering that the energy of an incident soliton should be equal to the sum of the energy of the transmitted and reflected solitons, that is to say assuming that no energy is lost through radiation, Cooney \emph{et al.} then suggested soliton specific reflection-transmission coefficients~\cite{Cooney1991,Lonngren1991} which appear supported by experimental data they obtained for IA solitons~\cite{Cooney1991}. The reflection of electrostatic IA solitons, and in particular the dependence of the reflection coefficient on different plasma properties, has in parallel been the object of many theoretical studies (see, \emph{e.~g.}, Refs.~\cite{Kuehl1983,Malik2007,Malik2008,Tomar2014}). In contrast, the analysis of the reflection of a MS soliton at an interface is to our knowledge limited to Nakata's theoretical prediction that the reflection of a MS soliton incident on a sharp density discontinuity can lead to a reflected soliton only if the discontinuity is a step up in density~\cite{Nakata1988}. 

The problem of energy~\cite{Fresnel1866} and momentum~\cite{Balazs1953,Jones1954} conservation when a traveling pulse is reflected at the interface between two media has a long and rich history. A fundamental question that is strongly rooted in this problem and which has attracted much attention is how to define the energy-momentum tensor and with it the momentum of an electromagnetic wave in a dielectric medium~\cite{Kemp2011}. Historically, the controversy has revolved around which of Minkowski's~\cite{Minkowski1908,Minkowski1910} or Abraham's~\cite{Abraham1909,Abraham1910} form is correct, though many alternative forms have been proposed since then by considering the particular basic properties of different media (see, \emph{e.~g.}, Refs.~\cite{Brevik1979,Kemp2011,Barnett2010,Dodin2012,Kemp2015}) including plasmas~\cite{Dewar1977,Novak1980}. While Minkowski's form seems to be the most natural in optics~\cite{Brevik2018}, there now appears to be a growing consensus that both forms are inherently valid and equivalent provided the total energy-momentum tensor is the same, and that how one divides the total energy-momentum tensor into light and matter components is arbitrary~\cite{Pfeifer2007,Pfeifer2009,Barnett2010a}. 

In this paper, we take on this problem by examining energy and momentum conservation as a MS soliton bounces back and forth in a vacuum-bounded magnetized plasma slab. Our goal with this work, however, is not to provide new results that would then favour one particular form for the energy-momentum tensor. Instead we focus on understanding the dynamics of energy and momentum in this system in light of the basic forces at play. To gain further insights into this complex dynamics, we use the simple analytical picture that solitons can provide to try to explain results from particle-in-cell simulations. This paper is organised as follows. In Sec.~\ref{Sec:config} we briefly recall the configuration of interest in this study. In Sec.~\ref{Sec:Soliton_scalings} we derive analytical formulae for the energy and momentum contents in our system assuming a magnetosonic (MS) soliton-like pulse. In Sec.~\ref{Sec:Simu_model} we describe the numerical model used to simulate this system and then analyse the reflection process through the EM fields dynamics at the interface. We then use simulation results to examine in detail energy and momentum conservation in Sec.~\ref{Sec:Energy} and \ref{Sec:Momentum}, respectively. Finally, we summarize our main findings in Sec~\ref{Sec:Summary}.

\section{Solitary pulse in a bounded magnetized plasma}
\label{Sec:config}

\begin{figure}
\begin{center}
\includegraphics[]{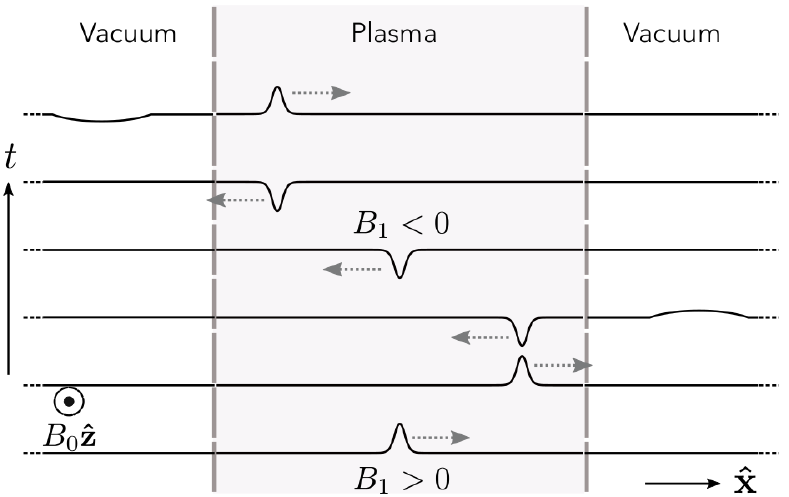}
\caption{Sketch of the plasma slab - pulse system. The profiles depict the time evolution of the magnetic field disturbance $B_1$, which changes sign upon reflection at the plasma-vacuum interface. Both the vacuum and the plasma slab are permeated by a constant and homogeneous background magnetic field $B_0 \mathbf{\hat{z}}$. The amplitude of the transmitted pulse is exaggerated for clarity.}
\label{Fig:Sketch}
\end{center}
\end{figure}

The focus in this work is on the propagation in a magnetized plasma of a magnetic perturbation perpendicularly to the background magnetic field, that is on perpendicular magnetosonic (also referred to as compressional-Alfv{\'e}n) waves~\cite{Swanson2003}. The configuration of particular interest here is illustrated in Fig.~\ref{Fig:Sketch}. It consists of a finite width plasma slab along $\mathbf{\hat{x}}$ immersed in a homogeneous background magnetic field $B_0 \mathbf{\hat{z}}$ in the presence of a solitary pulse 
\begin{equation}
\mathbf{B^{\pm}}(x) = [B_0 \pm B_1(x)]\mathbf{\hat{z}},
\end{equation}
with $B_1>0$ the magnetic perturbation propagating in the $\pm\mathbf{\hat{x}}$ direction. This configuration is analog to that studied previously by the author and co-workers~\cite{Gueroult2018b}, where it was indeed shown that the nature of the magnetic perturbation changes from compression ($B_1>0$) to rarefaction ($B_1<0$) and vice-versa upon reflection of the pulse at the plasma-vacuum interface. We further choose to limit ourselves to the overdense regime ($\omega_{pe}\gg\omega_{ce}$), and assume that the following ordering holds
\begin{equation}
\eta^2 = \frac{m_e}{m_i} \ll \frac{v_A}{c} = \eta\frac{\omega_{ce}}{\omega_{pe}}\ll \eta \ll \epsilon \ll 1.
\end{equation}
Here $m_e$ and $m_i$ are the electron and ion mass, $v_A = B_0/\sqrt{\mu_0n_0m_i}$ is the Alfv{\'e}n speed, $c$ is the speed of light, $\omega_{ce}=eB_0/m_e$ and $\omega_{pe}=[n_0e^2/(m_e\epsilon_0)]^{1/2}$ are the electron cyclotron and plasma frequency, respectively, $n_0$ is the background plasma density and $\epsilon = \max(B_1)/B_0$ is the normalized amplitude of the solitary pulse.

The notation adopted in this work to refer to the energy and longitudinal momentum contents of our system is summarised in Tab.~\ref{Tab:notation}. These quantities are, depending on the context, dressed with a $\hat{~}$ and indexed by a $\pm$ upperscript when referring to compression and rarefaction soliton-like pulses, respectively (see Sec.~\ref{Sec:Soliton_scalings}), and indexed with subscripts $p$ and $sp$ when referring to the pulse alone and the slab-pulse, respectively. All variables are properly defined in the text when introduced.

\begin{table}
\begin{center}
\caption{Notation for volume integrated energy and longitudinal momentum variables. Subscripts $p$ and $sp$ refer respectively to pulse and slab-pulse variables, while upper-scripts $\pm$ refer respectively to compression and rarefaction pulses. A $\hat{~}$ designates the analytical estimate obtained for a soliton-like pulse.}
\label{Tab:notation}
\begin{tabular} {c c c}
\hline
\hline
 & EM field & Mechanical\\
\hline
Energy & $\mathpzc{E}$ &  $\mathpzc{F}$ \\
Longitudinal momentum & $\mathpzc{P}$ &  $\mathpzc{Q}\,$ \\
\hline
\hline 
\end{tabular}
\end{center}
\end{table}

\section{Energy and momentum for a soliton-like pulse}
\label{Sec:Soliton_scalings}

To later facilitate the analysis of simulations results, we first derive in this section analytical formulas for the different energy and momentum contents of our slab-pulse system. 

To ease our calculations, we choose to consider the particular case of a magnetic perturbation \begin{equation}
B_1(x) =\epsilon B_0 \textrm{sech}^2[x\omega_{pe}\sqrt{\epsilon}/(2c)].
\end{equation}
The compression pulse $\mathbf{B^{+}}$ then matches the soliton solution for magnetosonic waves in an overdense plasma in the limit of a weakly nonlinear pulse $\epsilon\ll1$~\cite{Rau1998}. The longitudinal and transverse electric field $E_x$ and $E_y$ and the longitudinal and transverse electron and ion velocities $v_{e_x}$ and $v_{i_x}$ and $v_{e_y}$ and $v_{i_y}$ associated with this perturbation thus write~\cite{Gueroult2018b}
\begin{subequations}
\label{Eq:Soliton_scaling}
\begin{equation}
\frac{E_y^{+}}{v_AB_0} = \frac{v_{e_x}^{\pm}}{v_A} = \frac{v_{i_x}^{\pm}}{v_A} = \frac{n^{\pm}}{n_0}-1 = \epsilon \textrm{sech}^2\left[\frac{s\sqrt{\epsilon}}{2}\right],
\end{equation} 
\begin{equation}
-\frac{E_x^{+}}{v_AB_0} = \frac{v_{e_y}^{+}}{v_A} = \frac{\epsilon^{3/2}}{\eta} \textrm{sech}^2\left[\frac{s\sqrt{\epsilon}}{2}\right]\textrm{tanh}\left[\frac{s\sqrt{\epsilon}}{2}\right]
\end{equation}
and
\begin{equation}
v_{i_y}^{+}=0
\end{equation}
\end{subequations}
to lowest order in $\epsilon$ with $s=x\omega_{pe}/c$ the position normalized by the electron skin depth.

The rarefaction pulse $\mathbf{B^{-}}$ on the other hand is not a soliton as rarefaction solitons only exist for oblique propagation~\cite{Ohsawa2014}. Such a magnetic perturbation is thus not expected to preserve its shape as it propagates. We nevertheless define here for convenience and by analogy with the compression pulse the following fields, velocities and density perturbation for the rarefaction pulse
\begin{subequations}
\label{Eq:Soliton_scaling_m}
\begin{equation}
\frac{E_y^{-}}{v_AB_0} = \frac{v_{e_x}^{-}}{v_A} = \frac{v_{i_x}^{-}}{v_A} = 1-\frac{n^{-}}{n_0} = \epsilon \textrm{sech}^2\left[\frac{s\sqrt{\epsilon}}{2}\right],
\end{equation} 
\begin{equation}
-\frac{E_x^{-}}{v_AB_0} = \frac{v_{e_y}^{-}}{v_A} = - \frac{\epsilon^{3/2}}{\eta} \textrm{sech}^2\left[\frac{s\sqrt{\epsilon}}{2}\right]\textrm{tanh}\left[\frac{s\sqrt{\epsilon}}{2}\right]
\end{equation}
and
\begin{equation}
v_{i_y}^{-}=0,
\end{equation}
\end{subequations}
and keep in mind that these quantities will evolve in time.

It is worth noting here that since $v_{\alpha_x}>0$ ($\alpha=e,i$) both for compression and rarefaction pulses,  and because $v_{i_x}^{\pm}\gg v_{i_y}^{\pm}$, the kinetic energy flux is about $n_0 m_i {v_{i_x}}^3/2$ and thus always in the $\mathbf{\hat{x}}$ direction. This is also true for the EM energy flux since, as $E_y^{+} = E_y^{-}$, the Poynting vector $\bm{\Pi} = \mu_0^{-1}\mathbf{E}\times\mathbf{B}$ is such that $\bm{\Pi}\cdot\mathbf{\hat{x}}>0$ irrespective of the nature of the pulse. For the rarefaction pulse propagating towards $-\mathbf{\hat{x}}$ shown in Fig.~\ref{Fig:Sketch}, the energy flow is thus opposed to the direction of propagation. This is characteristic of backward waves. 

\subsection{Energy}

For the overdense regime considered here $v_A\ll c$ which in turn guarantees that the electromagnetic (EM) energy is to lowest order equal to the magnetic field energy. Considering the pulse alone, the EM energy in the pulse is hence
\begin{align}
\hat{\mathpzc{E}}_p^{\pm} & \doteq \frac{1}{2\mu_0}\int_{-\infty}^{\infty} [\pm B_1]^2dx\nonumber\\
 & = \frac{4c{B_0}^2\epsilon^{3/2}}{3\mu_0\omega_{pe}},
\label{Eq:EM_energy_pulse}
\end{align}
where we used the change of variable $\chi = x\omega_{pe}\sqrt{\epsilon}/(2c)$ to carry out the integral. Clearly, $\hat{\mathpzc{E}}_p^{+}=\hat{\mathpzc{E}}_p^{-}=\hat{\mathpzc{E}}_p$ so that the pulse EM energy is, to lowest order in $v_A/c$, the same for compression and rarefaction pulses. The ordering $v_A\ll c$ also ensures that the kinetic energy is to lowest order equal to the ion kinetic energy along $\mathbf{\hat{x}}$~\cite{Gueroult2018b} and 
\begin{align}
\hat{\mathpzc{F}}_p^{\pm} & \doteq\frac{1}{2} \int_{-\infty}^{\infty} n_0 m_p {v_{i_x}^{\pm}}^2(x) dx\nonumber\\
 & = \frac{4n_0 m_i{v_A}^2\epsilon^{3/2}c}{3\omega_{pe}}.
 \label{Eq:K_energy_pulse}
\end{align}
This shows that the kinetic energy is also identical for compression and rarefaction pulses to lowest order in $\epsilon$ and $\hat{\mathpzc{F}}_p^{+}=\hat{\mathpzc{F}}_p^{-}=\hat{\mathpzc{F}}_p$. In addition, since $v_A = {B_0}/\sqrt{\mu_0n_0m_i}$, $\hat{\mathpzc{E}}_p = \hat{\mathpzc{F}}_p$.

Considering now the slab-pulse system, its EM energy writes
\begin{equation}
\hat{\mathpzc{E}}_{sp}^{\pm} \doteq \int_{-L/2}^{L/2} \frac{B^{\pm}(x)^2}{2\mu_0} dx,
\label{Eq:slab_pulse_magnetic_energy}
\end{equation}
with $L$ the length of the simulation domain. One immediately finds that $\hat{\mathpzc{E}}_{sp}^{+}\geq \hat{\mathpzc{E}}_{sp}^{-}$, that is to say that the EM energy in the system is greater in the case of a compression pulse than for a rarefaction pulse. Quantitatively, one gets 
\begin{equation}
\hat{\mathpzc{E}}_{sp}^{+} = \hat{\mathpzc{E}}_{sp}^{-} + 8 \sqrt{\epsilon} \frac{c}{\omega_{pe}}\frac{{B_0}^2}{\mu_0}
\label{Eq:diff}
\end{equation}
to lowest order in $\epsilon$, where it has been assumed that the soliton width $w_p\sim c/(\sqrt{\epsilon}\omega_{pe})$ is small compared to the plasma slab width $L_p < L$. Under a cold plasma assumption, the kinetic energy of the slab-pulse system $\hat{\mathpzc{F}}_{sp}^{\pm}$ is simply $\hat{\mathpzc{F}}_p^{\pm}$.

\subsection{Momentum}

As mentioned in the introduction, one is faced with having to choose between different options when defining the momentum of the slab-pulse system, and this choice has to do with how to partition momentum in the system between field and matter. Historically two main frameworks have been used to study the momentum of light propagating through a linear medium, relying respectively on Abraham's and Minkowski's definition for the electromagnetic wave momentum density $\mathbf{g}^A = \mathbf{E}\times\mathbf{H}/c^2$~\cite{Abraham1909,Abraham1910} and $\mathbf{g}^M = \mathbf{D}\times\mathbf{B}$~\cite{Minkowski1908,Minkowski1910}. Here $\mathbf{D}$ and $\mathbf{H}$ are the electric displacement field and the magnetic field strength, respectively. 

Instead of adopting right-away a particular framework, we choose in this work to examine separately the free space electromagnetic wave momentum density $\epsilon_0\mathbf{E}\times\mathbf{B}$ and the mechanical momentum density $\rho\mathbf{v}$, and relate our findings to Abraham's and Minkowski's when appropriate. This approach has the advantage as we will see, that it clearly shows momentum exchange between the macroscopic electromagnetic field and the plasma~\cite{Frias2012}. To gain additional insights, we further choose to examine separately the EM momentum of the pulse and the EM momentum of the slab-pulse system, as indicated by the $s$ and $sp$ subscripts, respectively. 

Note that since the relative permeability $\mu_r$ in a plasma is close to unity, our definition of EM field momentum matches Abraham's momentum $\mathbf{g}^A$. Also, in a cold magnetized plasma the susceptibility tensor~\cite{Swanson2003}
\begin{equation}
\bm{\chi}(\omega) = \begin{pmatrix}
\chi_{\perp} & -j\chi_{\times} & 0\\\
j\chi_{\times} & \chi_{\perp} & 0\\
0 & 0 & \chi_{\parallel}
\end{pmatrix}
\end{equation}
is non-diagonal so that the constitutive relation $\mathbf{D}= \epsilon_0(\mathds{1}+\bm{\chi})\mathbf{E}$ is in general tensorial and, as a result, the relation between Abraham's and Minkowski's momenta is not immediate. However, for low frequency waves $\chi_{\perp}(\omega)\gg\chi_{\times}(\omega)$, and one can then write $D_y\sim \epsilon_0(1+\chi_{\perp})E_y$ with $\chi_{\perp}\sim(c/{v_A})^2$. In this case Abraham's and Minkowski's definitions for the electromagnetic wave momentum density $g_x$ perpendicular to both the plasma-vacuum interface and the background magnetic field differ by a factor $\epsilon_{\perp}= 1+\chi_{\perp}$. For an overdense plasma $v_A/c\ll1$ and then one simply gets $\epsilon_{\perp}\sim(c/{v_A})^2$.

With these definitions in hand, the pulse's electromagnetic linear momentum density in the $\mathbf{\hat{x}}$ direction then writes
\begin{equation}
g_{x}^{\pm}(x)  \doteq \pm\epsilon_0 E_y(x) B_1(x).
\end{equation}
From there the EM momentum along $\mathbf{\hat{x}}$ associated with the pulse is 
\begin{equation}
\hat{\mathpzc{P}}_p^{\pm} \doteq \int_{-\infty}^{\infty} g_{x}^{\pm}(x) dx = {\pm}\frac{8 v_A{B_0}^2\epsilon^{3/2}}{3\mu_0c\omega_{pe}}.
\label{Eq:momentum_EM}
\end{equation}
Similarly, the linear mechanical momentum density is to lowest order in $\epsilon$ (that is neglecting the contribution of the density perturbation),
\begin{equation}
\label{Eq:mecha_momentum_density}
p_{x}^{\pm}(x)  \doteq n_0 m_p {v_{i_x}}(x) = \epsilon n_0m_p {v_A} \textrm{sech}^2\left[x\frac{\omega_{pe}\sqrt{\epsilon}}{2c}\right],
\end{equation}
so that
\begin{equation}
\hat{\mathpzc{Q}\,}_p^{\pm}  \doteq \int_{-\infty}^{\infty} p_{x}^{\pm}(x) dx = \frac{4v_A n_0 m_pc\sqrt{\epsilon}}{\omega_{pe}}.
\label{Eq:momentum_Mech}
\end{equation}
The pulse EM momentum $\hat{\mathpzc{P}}_p^{\pm}$ then depends on the nature of the pulse, while the pulse mechanical momentum $\hat{\mathpzc{Q}\,}_p^{\pm}$ does not. Furthermore, one finds
\begin{equation}
\frac{|\hat{\mathpzc{P}}_p^{\pm}|}{\hat{\mathpzc{Q}\,}_p^{\pm}} = \frac{2}{3}\left(\frac{v_A}{c}\right)^2\epsilon \ll 1
\label{Eq:momentum_ratio}
\end{equation}
so that the EM pulse's momentum is small compared to the mechanical momentum. 

Looking now at the the slab-pulse system, for a stationary background plasma the slab-pulse mechanical momentum $\hat{\mathpzc{Q}\,}_{sp}^{\pm}$ is equal to that of the pulse alone $\hat{\mathpzc{Q}\,}_p^{\pm}$. It is hence also independent of the pulse nature. Meanwhile,
the EM longitudinal momentum linear density is to lowest order in $\epsilon$ simply $\epsilon_0 E_y(x) B_0$ and the EM momentum along $\mathbf{\hat{x}}$ is hence
\begin{align}
\hat{\mathpzc{P}}_{sp}^{\pm}  \doteq &\epsilon_0 B_0\int_{-\infty}^{\infty} E_y(x) dx \nonumber\\
= & \frac{4v_A{B_0}^2\sqrt{\epsilon}}{\mu_0 c\omega_{pe}}\nonumber\\
= &\left(\frac{v_A}{c}\right)^2 \hat{\mathpzc{Q}\,}_p^{\pm}.
\label{Eq:momentum_EM_tot}
\end{align}
This means that, unlike the pulse EM momentum $\hat{\mathpzc{P}}_p^{\pm}$, the slab-pulse EM momentum $\hat{\mathpzc{P}}_{sp}^{\pm}$ does not depend on the nature of the pulse. 

The free space electromagnetic momentum of the slab pulse system is thus $\chi_{\perp}$ smaller than its mechanical counterpart. Summing these two contributions one gets 
\begin{equation}
\hat{\mathpzc{P}}_{sp}^{\pm}+\hat{\mathpzc{Q}\,}_{sp}^{\pm} = \epsilon_0\epsilon_{\perp}\hat{\mathpzc{P}}_{sp}^{\pm}
\end{equation}
which matches Minkowski's momentum. This is consistent with Nakamura's derivation for an MHD wave~\cite{Nakamura2016}. Note that both of these contributions are in the positive $\mathbf{\hat{x}}$ irrespective of the nature of the pulse. Note also that the next order correction to $\hat{\mathpzc{P}}_{sp}^{\pm}$, which arises from the magnetic perturbation, happens to be $\hat{\mathpzc{P}}_p^{\pm}$. At this higher order the EM momentum then depends on the nature of the pulse and is larger for a compression pulse than for a rarefaction pulse. 

The main results from this section, which will be useful for the rest of the discussion, are summarised in Tab.~\ref{Tab:soliton_var}. Since, as we have shown, the pulse mechanical and EM energy are equal and independent of the nature of the pulse, we drop the $^{\pm}$ upper-script and simply write it as $\hat{\mathpzc{E}}_p = \hat{\mathpzc{F}}_p$ in the rest of this study. Similarly, since the mechanical momentum for the pulse and the pulse EM are identical and independent of the nature of the pulse, we simply write $\hat{\mathpzc{Q}\,} = \hat{\mathpzc{Q}\,}_p^{\pm} = \hat{\mathpzc{Q}\,}_{sp}^{\pm}$.

\begin{table}
\begin{center}
\caption{Volume integrated energy and longitudinal momentum for a soliton like pulse. Definitions of energies and momenta are given in Eqs.~(\ref{Eq:EM_energy_pulse}) and (\ref{Eq:K_energy_pulse}) and Eqs.~(\ref{Eq:momentum_EM}), (\ref{Eq:momentum_EM_tot}) and (\ref{Eq:momentum_Mech}), respectively. }
\label{Tab:soliton_var}
\begin{tabular} {c c}
\hline
\hline
\multicolumn{2}{c}{Energy}\\
\hline
pulse EM & $\hat{\mathpzc{E}}_p= \frac{\displaystyle 4c{B_0}^2\epsilon^{3/2}}{\displaystyle 3\mu_0\omega_{pe}}$  \\
mechanical & $\hat{\mathpzc{F}}_p=\hat{\mathpzc{E}}_p$\\
\hline
\multicolumn{2}{c}{Longitudinal momentum}\\
\hline
pulse EM & $\hat{\mathpzc{P}_p}^{\pm} = {\pm}\frac{\displaystyle 8 v_A{B_0}^2\epsilon^{3/2}}{\displaystyle 3\mu_0c\omega_{pe}} = 2 \epsilon \hat{\mathpzc{P}}_{sp}/3$  \\
pulse-slab EM & $\hat{\mathpzc{P}}_{sp} = \frac{\displaystyle 4v_A{B_0}^2\sqrt{\epsilon}}{\displaystyle \mu_0 c\omega_{pe}} = \left(\frac{\displaystyle v_A}{\displaystyle c}\right)^2\hat{\mathpzc{Q}\,}$  \\
mechanical & $\hat{\mathpzc{Q}\,} = \frac{\displaystyle 4v_A n_0 m_pc\sqrt{\epsilon}}{\displaystyle \omega_{pe}}$ \\
\hline
\hline 
\end{tabular}
\end{center}
\end{table}

\section{Numerical simulations}
\label{Sec:Simu_model}

\subsection{Numerical model}

The numerical data presented and discussed in this study are particle-in-cell simulation results produced with the 1d version of the Epoch code~\cite{Arber2015} modified to allow for a background magnetic field. 

The slab model considered is analogous to that previously studied by the author and co-workers~\cite{Gueroult2018b}. Briefly, it consists of a 1d simulation domain of length $L$ along $\mathbf{\hat{x}}$. Centered in this domain is initially a fully ionised hydrogen plasma slab of length $L_p<L$, and the entire simulation domain (plasma plus surrounding vacuum) is permeated by a background uniform magnetic field $\mathbf{B} = B_0\mathbf{\hat{z}}$. At $t=0$, a compression soliton propagating towards $\mathbf{\hat{x}}$ is initialized in the middle of the plasma slab.

Simulations parameters are given in Tables~\ref{Tab:input_deck} and \ref{Tab:dimensionless}. The choice of this particular set of parameters is motivated by PIC requirements. Indeed, since simulation time scales as $\eta^{-1}/(\sqrt{\epsilon}\omega_{ce})$ (pulse speed and width proportional to $v_A$ and $c/[\sqrt{\epsilon}\omega_{pe}]$, respectively), simulations of soliton dynamics in PIC calls for strong magnetic fields. Additionally, since the soliton width to Debye length ratio depends only on the pulse amplitude and plasma temperature but not on the density (or the magnetic field), an increase in plasma density does not necessarily lead to a greater computational cost. Simulations of soliton dynamics in over-dense regimes ($\omega_{pe}/\omega_{ce}\gg1$) are thus best done in PIC at high plasma density. Yet, it is worth insisting that this is purely a consequence of the requirement of PIC models, and that the findings presented here are hence expected to hold at lower densities and magnetic field as long as $\omega_{pe}/\omega_{ce}\gg1$.

\begin{figure}
\begin{center}
\includegraphics[]{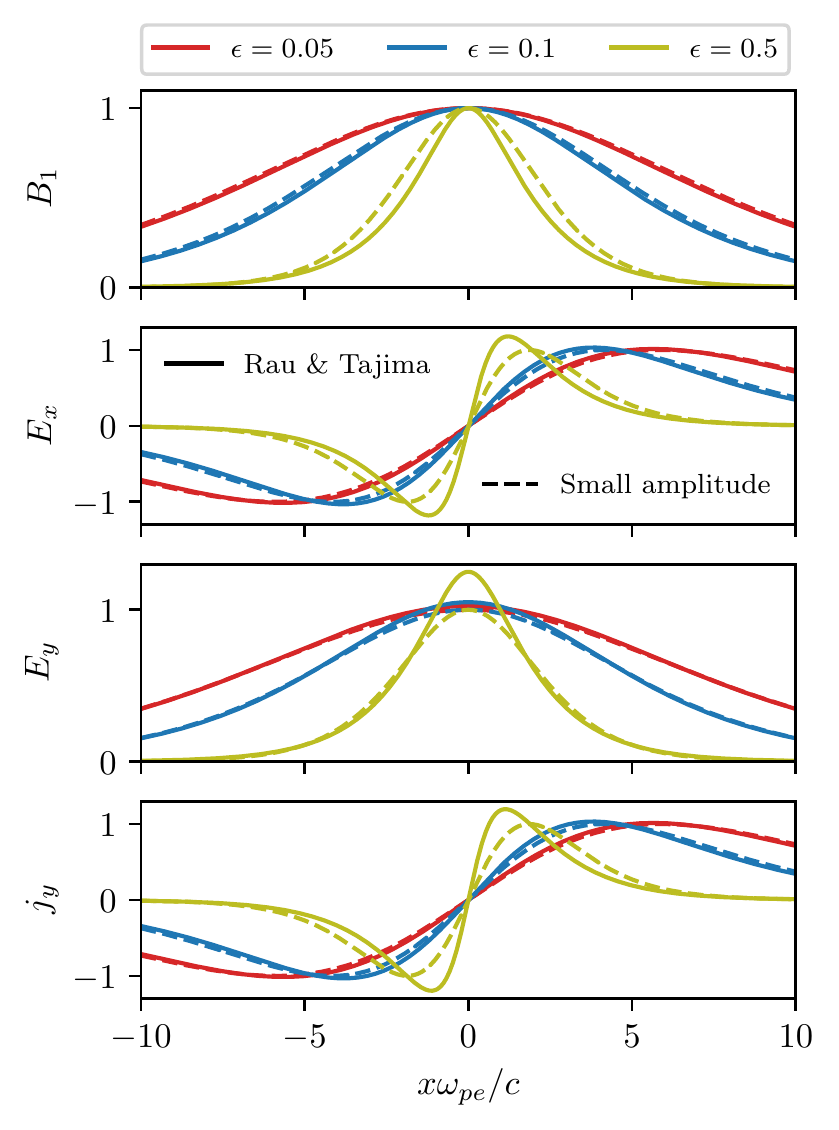}
\caption{Profiles of solitary wave solutions for the analytical small-amplitude sech$^2(s\sqrt{\epsilon}/2)$ magnetic perturbation given in Eqs.~(\ref{Eq:Soliton_scaling}) (dashed line) and for the large amplitude numerically integrated solitary wave solution from Rau and Tajima~\cite{Rau1998} (solid line) for three pulse amplitudes $\epsilon = 0.05, 0.1$ and $0.5$. All quantities are normalized by the peak value of the small-amplitude solution for the corresponding $\epsilon$. Differences become noticeable for $\epsilon\geq0.1$. Background plasma parameters are identical to those used in the simulation and given in Table~\ref{Tab:input_deck}. }
\label{Fig:RauTajimaVsSmallAmpSoliton}
\end{center}
\end{figure}

Besides the use in this work of a physical electron to proton mass ratio $\eta^2$, another noteworthy difference compared to Ref.~\cite{Gueroult2018b} is that the initial perturbation used in this work does not use the small amplitude analytical solution given in Eqs.~(\ref{Eq:Soliton_scaling}) but instead uses the numerical soliton solution valid for larger amplitude solitary waves derived by Rau and Tajima~\cite{Rau1998}. While, as illustrated in Fig.~\ref{Fig:RauTajimaVsSmallAmpSoliton}, the difference between these two solutions remains small for $\epsilon=0.1$, it was found that intializing the simulation with the small amplitude solution Eqs.~(\ref{Eq:Soliton_scaling}) leads initially to the propagation of electromagnetic waves before the perturbation reaches a shape-preserving profile. On the other hand, this behaviour was not observed when using the solution for larger amplitude waves.  This difference and the presence of these supplemental waves in the first instants of the simulation can be understood in light of the fact that Ampere's law
\begin{equation}
\frac{\partial E_y}{\partial t} = -c^2 \frac{\partial B_z}{\partial x}-\frac{j_y}{\epsilon_0}
\end{equation}
is only satisfied by the solution in Eq.~(\ref{Eq:Soliton_scaling}) in the limit of ${v_A}^2/c^2\rightarrow 0$ and $\epsilon\ll1$, while it is built-in the solution of Ref.~\cite{Rau1998}. Initialising the soliton from Rau and Tajima's solution thus ensures minimal spurious effects.

\begin{table}
\begin{center}
\caption{Input parameters for particle-in-cell simulations. Background density $n_0$, background magnetic field $B_0$, electron and ion temperature $T_e$ and $T_i$ and pulse amplitude $\epsilon$.}
\label{Tab:input_deck}
\begin{tabular}{c c}
\hline
\hline
\multicolumn{2}{c}{Plasma parameters}\\
\hline
$n_0$~[$\times 10^{21}$~m$^{-3}$] & $2$\\
$B_0$~[T] & $1.434$\\
$T_e = T_i$~[eV] & $3~10^{-2}$\\
$\epsilon = \delta B/B_0$ & $0.1$\\
$\eta^{-2} = m_i/m_e$ & 1836\\
\hline
\multicolumn{2}{c}{Simulations parameters}\\
\hline
Grid size  & $\lambda_D$\\
Particles per species & $9\times 10^{6}$\\
Plasma region & $[0.1,0.9]L$ \\
\hline
\hline
\end{tabular}
\end{center}
\end{table}

\begin{table}
\begin{center}
\caption{Dimensionless parameters derived from simulations inputs. $v_{the}$ and $v_{thi}$ are the electron and ion thermal speeds, $\rho_{e,th}$ and $\rho_{i,th}$ are the electron and ion thermal Larmor radii, $w_p = c/(\sqrt{\epsilon}\omega_{pe})$ is the pulse width and $\Xi = (eE_x)^2/(2m_p{\omega_{ci}}^2)$ is the ion ponderomotive energy.  }
\label{Tab:dimensionless}
\begin{tabular}{c c c }
\hline
\hline
Frequencies & $\omega_{ce}/\omega_{pe}$ & $0.1$\\
\hline
\multirow{3}{*}{Speeds} & $v_A/c$ & $2.3~10^{-3}$\\
 & $v_A/v_{the}$ & 10\\
& $v_A/v_{thi}$ & 410\\
\hline
\multirow{6}{*}{Lengths} & $L\omega_{pe}/w_p$ & $31$\\
 & $L\omega_{pe}/c$ & $97$\\
 & $L/\lambda_D$  & $400320$\\
 & $L_p\omega_{pe}/c$ & $78$\\
 & $\rho_{e,th}/w_p$ & $ 7.7~10^{-4}$\\
 & $\rho_{i,th}/w_p$ & $ 3.3~10^{-2}$\\
\hline
Energies & $k_bT_i/\Xi$ & $4.3~10^{-5}$\\
\hline
\hline
\end{tabular}
\end{center}
\end{table}

\subsection{Global results}

\begin{figure}
\begin{center}
\includegraphics[]{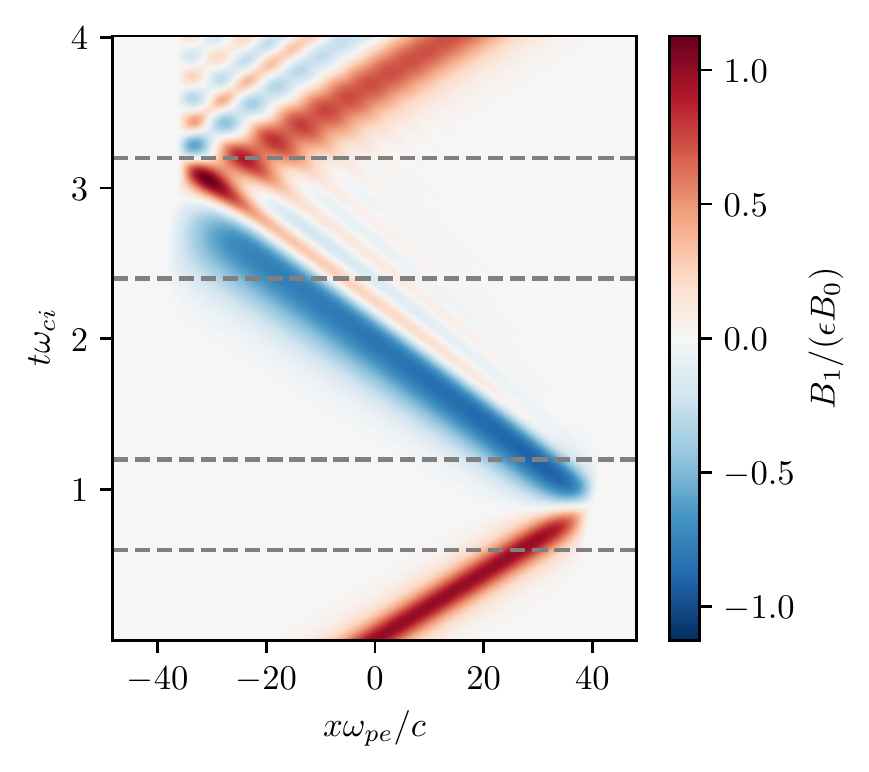}
\caption{Map of the magnetic field perturbation $B_1$ corresponding at $t=0$ to a magneto-sonic soliton of amplitude $\epsilon B_0$ at $x = 0$ in a finite width plasma slab. The horizontal dashed lines approximately indicate the start and finish of reflection events. }
\label{Fig:BMap}
\end{center}
\end{figure}

The transformation of the magnetic perturbation from compression to rarefaction and vice-versa upon reflection at the plasma-vacuum interface previously uncovered in Ref.~\cite{Gueroult2018b} is confirmed in these new simulations. This is immediately seen in Fig.~\ref{Fig:BMap}. This same figure further underlines the profile-preserving nature of the original compression soliton up until the first reflection ($t\omega_{ci}\sim0.9$). Past this point results  show that the pulse is no longer a soliton as one observes a clear change in pulse profile over time after the first reflection. This result seems consistent with Nakata's prediction that a reflection can only produce a soliton when density goes up across the discontinuity~\cite{Nakata1988}.  

\subsection{Reflection dynamics}

\begin{figure}
\begin{center}
\includegraphics[]{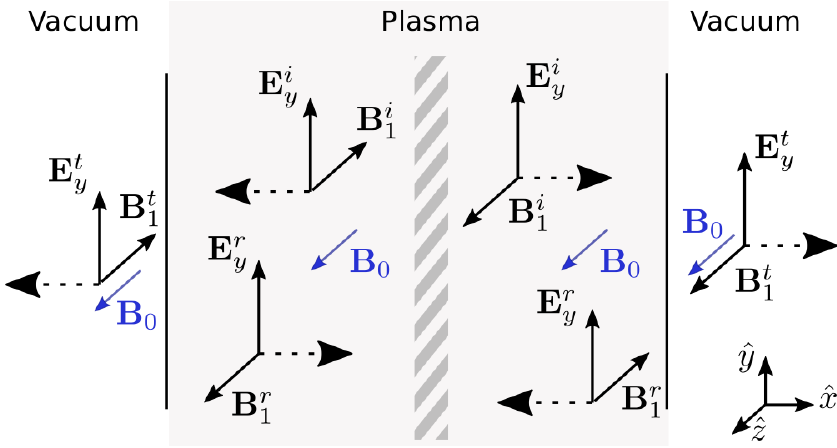}
\caption{Electromagnetic field convention for incident, reflected and transmitted pulses at the plasma-vacuum interface.  }
\label{Fig:ReflectionConfig}
\end{center}
\end{figure}

\begin{figure*}
\begin{center}
\subfigure[]{\includegraphics[height=7cm]{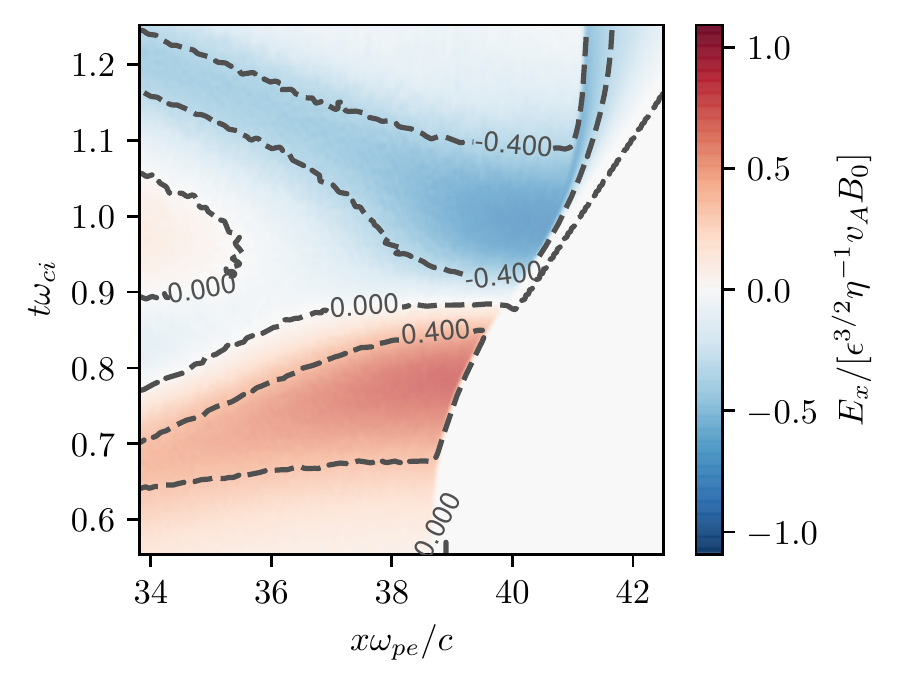}\label{Fig:Ex_int}}\subfigure[]{\includegraphics[height=7cm]{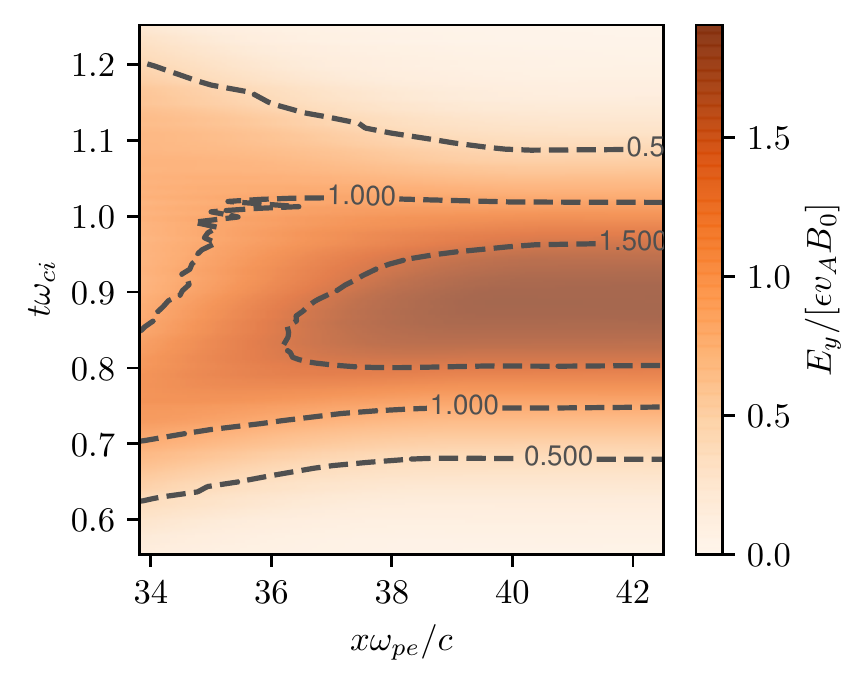}\label{Fig:Ey_int}}
\subfigure[]{\includegraphics[height=7cm]{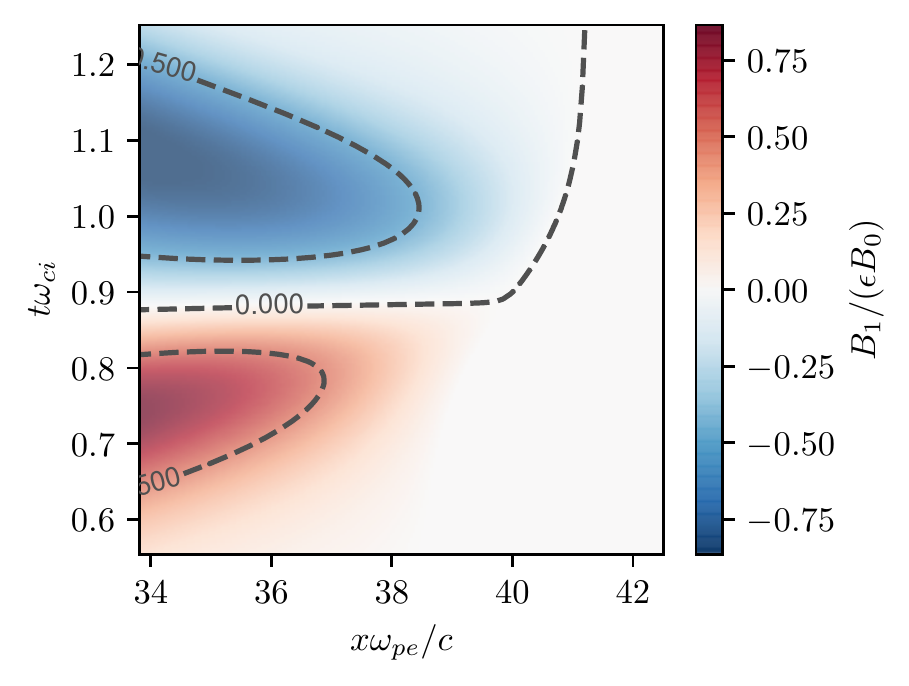}\label{Fig:Bz_int}}\subfigure[]{\includegraphics[height=7cm]{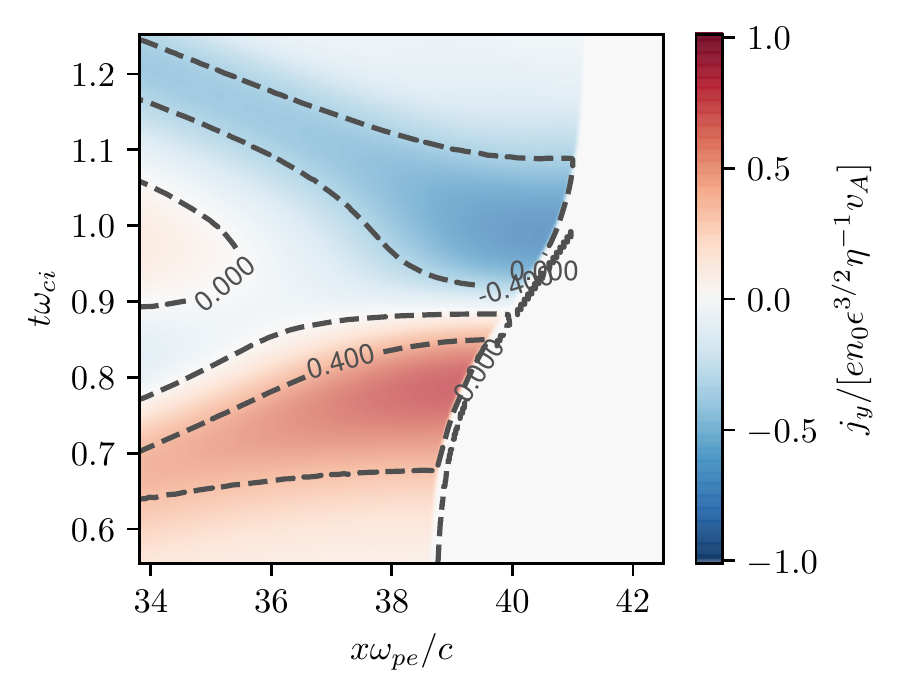}\label{Fig:Jy_int}}
\subfigure[]{\includegraphics[height=7cm]{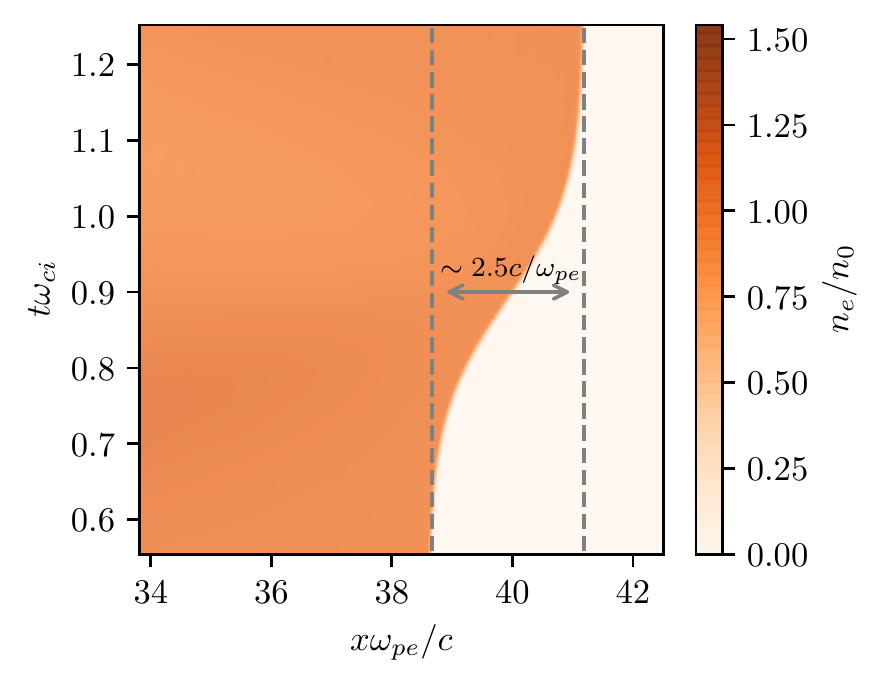}\label{Fig:ne_int}}
\caption{Map of the normalized longitudinal electric field $E_x/[\epsilon^{3/2}\eta^{-1}v_AB_0]$ \subref{Fig:Ex_int}, transverse electric field  $E_y/[\epsilon v_AB_0]$ \subref{Fig:Ey_int}, magnetic perturbation $B_1/[\epsilon B_0]$  \subref{Fig:Bz_int}, transverse current $j_y/[e n_0\epsilon^{3/2}\eta^{-1}v_A]$ \subref{Fig:Jy_int} and electron density $n_e/n_0$ \subref{Fig:ne_int} near the plasma-vacuum interface during the first reflection. }
\label{Fig:FieldsMapRef}
\end{center}
\end{figure*}

Before studying in detail in Sec.~\ref{Sec:Energy} and \ref{Sec:Momentum} the effect of reflection at the interface on energy and momentum in the system, it is useful to examine how the pulse reflection itself takes place.  To this end we use the convention for incident, reflected and transmitted pulses depicted in Fig.~\ref{Fig:ReflectionConfig}, consistent with that already used in Ref.~\cite{Gueroult2018b}, and examine the dynamics at the interface as shown in Fig.~\ref{Fig:FieldsMapRef}. 

Let us first introduce 
\begin{subequations}
\begin{equation}
r_F \doteq \frac{B_1^r}{B_1^i} = \frac{\tilde{\kappa}^{1/2}-1}{\tilde{\kappa}^{1/2}+1}
\label{Eq:magnetic_reflection_coeff}
\end{equation}
and
\begin{equation}
t_F \doteq \frac{B_1^t}{B_1^i} = \frac{2}{\tilde{\kappa}^{1/2}+1}
\label{Eq:magnetic_transmission_coeff}
\end{equation}
\end{subequations}
the Fresnel reflection and transmission coefficients for the magnetic perturbation at the plasma-vacuum interface, where the complex dielectric constant $\tilde{\kappa}$ is related to the wave index $\tilde{\eta}$ through $\tilde{\eta} = \sqrt{\mu_0/(\tilde{\kappa}\epsilon_0)}$. For a low-frequency extraordinary wave in the overdense regime $\tilde{\kappa}^{1/2}\sim\omega_{pi}/\omega_{ci}\gg 1$~\cite{Swanson2003}. One then gets $r_F=1-2\omega_{ci}/\omega_{pi} = 1-2v_A/c$ and $t_F= \omega_{ci}/(2\omega_{pi}) = 2v_A/c$ to lowest order in $\omega_{ci}/\omega_{pi}$, which implies a quasi perfect reflection of the pulse. Going back to the field convention shown in Fig.~\ref{Fig:ReflectionConfig}, this translates into incident and reflected magnetic perturbations of nearly equal amplitude but opposite signs, and as a result to a total magnetic field in the plasma at the interface $B_1^i-B_1^r\ll B_1^i$ that is consistent with weak transmission. On the other hand, since the sign of the transverse electric field of both the incident and reflected pulses is the same, Eq.~(\ref{Eq:magnetic_reflection_coeff}) implies that $E_y$ at the interface is about twice the amplitude of the incident pulse $E_y^i+E_y^r\sim 2E_y^i$. As it will become clear, these two simple observations allow interpreting many features of the soliton reflection.

Consider the case of an incident compression magnetic perturbation ($B_1>0, E_y>0$). Since $v_{e_x}\sim E_y/B_z$, the local increase in $E_y$ at the interface leads to an increase in longitudinal electron velocity. In response to this change in $v_{e_x}$, the longitudinal electric field $E_x>0$ at the interface must then increase to accelerate ions so as to maintain a nearly zero longitudinal current (Fig.~\ref{Fig:Ex_int}). Concurrently, since the transverse current $j_y$ is mostly carried by the cross-field electron drift $v_{e_y}\sim-E_x/B_z$,  this increase in $E_x$ also implies that the transverse current $j_y\sim e n_e E_x$ grows near the interface (Fig.~\ref{Fig:Jy_int}). This behaviour proceeds up until the incident pulse reaches its maximum value (for $t\omega_{ci}\sim0.8$), at which point $E_y$ at the interface begins to decrease (Fig.~\ref{Fig:Ey_int}) and the entire process reverts. The progressive decrease in $v_{e_x}$ leads to a decrease in $E_x$ which eventually changes sign as ions have now overpassed electrons. As a consequence of this longitudinal field reversal, $j_y$ becomes negative. Since $\partial B_1/\partial x = -\mu_0 j_y$, this negative transverse current density is consistent with a negative magnetic perturbation emerging from reflection at the interface.

Quantitatively, assuming that $E_y$ at the interface is twice the transverse electric field of the incident pulse at anytime implies that $v_{e_x}$ at the interface is twice the electron longitudinal velocity in the pulse. From Eq.~(\ref{Eq:Soliton_scaling}) this would yield a displacement of the interface
\begin{align}
\Delta x & \doteq \int_{-\infty}^{\infty} 2\epsilon v_A \textrm{sech}^2\left[\frac{\omega_{pe}\sqrt{\epsilon}v_A}{2c}t\right]dt\nonumber\\
 & = 8\sqrt{\epsilon}\frac{c}{\omega_{pe}}.
\end{align} 
For $\epsilon=0.1$ this gives $\Delta x\sim2.5 c/\omega_{pe}$, which matches nearly exactly the plasma expansion observed during the reflection of a compression pulse into a rarefaction pulse in simulations as highlighted in Fig.~\ref{Fig:ne_int}. 

If we now consider a negative magnetic perturbation propagating along $-\mathbf{\hat{x}}$, we find that the same picture holds since $E_y>0$ and $v_{e_x}>0$ in both cases. This incidentally supports that the reflection of a rarefaction pulse propagating towards $-\mathbf{\hat{x}}$ leads to a displacement of the plasma-vacuum interface towards $\mathbf{\hat{x}}$, as already observed in Ref.~\cite{Gueroult2018b}

\section{Energy}
\label{Sec:Energy}
\subsection{High- and low-energy state of the slab-pulse system}

Consider first the EM energy of the pulse-slab system 
\begin{equation}
\mathpzc{E}_{sp} \doteq \frac{1}{2}\int_V \left[\epsilon_0 |\mathbf{E}|^2+{\mu_0}^{-1}|\mathbf{B}|^2\right]dV,
\end{equation}
with $V$ the simulation domain. This is the numerical analog of $\hat{\mathpzc{E}}_{sp}^{\pm}$ obtained in Eq.~(\ref{Eq:slab_pulse_magnetic_energy}) for a soliton-like pulse. The time evolution of $\mathpzc{E}_{sp}$ is governed by Poynting's theorem with
\begin{equation}
\label{Eq:Poynting}
\frac{d\mathpzc{E}_{sp}}{dt} = -\oiint_{\partial V}\bm{\Pi}\cdot{\mathbf{dA}}-\int_{V} \bm{j}\cdot\mathbf{E}~dV,
\end{equation}
where 
$\bm{\Pi}\doteq{\mu_0}^{-1}\mathbf{E}\times\mathbf{B}$ is the Poynting vector and $\bm{j}$ is the current density. 

\begin{figure}
\begin{center}
\includegraphics[]{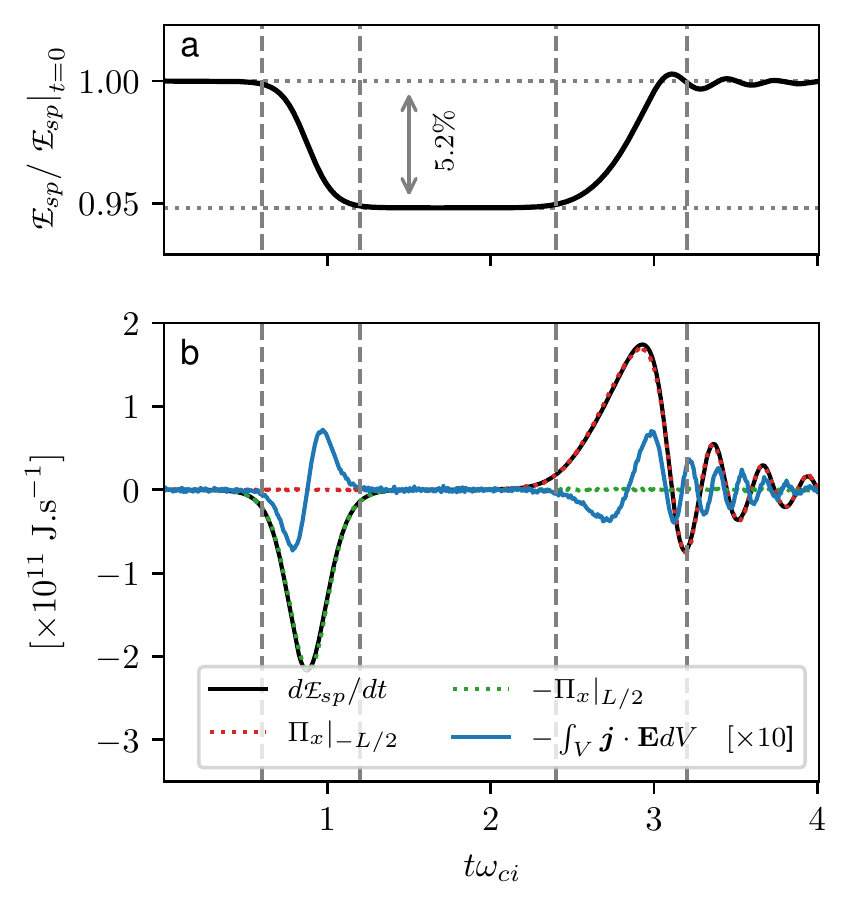}
\caption{Time evolution of the EM energy of the pulse-slab system $\mathpzc{E}_{sp}$ (a) and breakdown of the contribution of Joule losses and flux of the Poynting vector through the axial ends of the simulation domain (b). Joule losses plotted in blue in the lower panel are multiplied by $10$ for clarity. The vertical dashed gray lines depict the start and finish of reflection events, consistent with those shown in Fig.~\ref{Fig:BMap}. }
\label{Fig:Energy_tot}
\end{center}
\end{figure}

Away from reflection events, that is to say when the pulse is far from plasma-vacuum interfaces, one expects EM fields in the vacuum region to be null. As a result the contribution of the Poynting flux on the right hand side of Eq.~(\ref{Eq:Poynting}) should be zero. In addition, far away from reflections one can simply consider a pulse in an infinite plasma. For a soliton this would then lead to zero Joule losses. To see this, consider first that soliton profiles given in Eqs.~(\ref{Eq:Soliton_scaling}) imply $j_x = 0$ and $j_y\sim en_e v_{e_y}$. Now since the transverse electron velocity $v_{e_y}$ and the transverse electric field $E_y$ are respectively antisymmetric and symmetric with respect to the center of the pulse~\cite{Gueroult2018b}, this yields that the volume integral of $\bm{j}\cdot\mathbf{E}\sim j_y E_y$ is indeed zero. Putting these pieces together, Eq.~(\ref{Eq:Poynting}) then predicts no change in EM energy $\mathpzc{E}_{sp}$ away from reflection events. As shown in Fig.~\ref{Fig:Energy_tot}.a, this prediction is well verified in simulations with $\mathpzc{E}_{sp}$ constant in this case to less than $1\permil$. This consistency with soliton theory is particularly remarkable considering that, as already underlined, only the initial pulse up to the first reflection is a soliton.

\begin{figure}
\begin{center}
\includegraphics[]{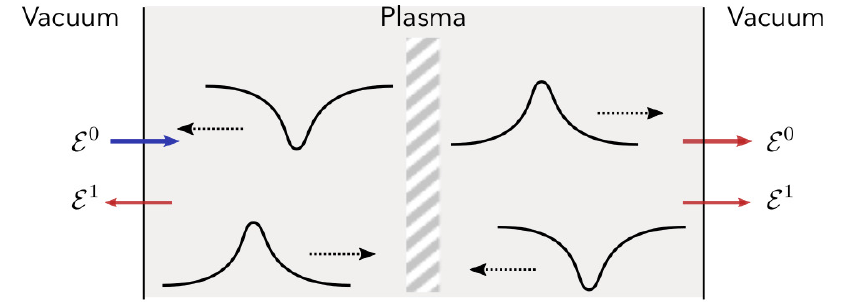}
\caption{Energy exchange for the slab-pulse system upon transformation from a rarefaction (compression) to a compression (rarefaction) pulse. $\mathcal{E}^0$ is the energy exchange corresponding to the change in pulse nature assuming perfect reflection. $\mathcal{E}^1\ll\mathcal{E}^0$ is the pulse energy loss due to the partial transmission of the magnetic field pertubation upon reflection.  }
\label{Fig:Energy}
\end{center}
\end{figure}

Looking now at reflections and assuming for now perfect ($r_F=1$) reflection, Eq.~(\ref{Eq:slab_pulse_magnetic_energy}) predicts a relative decrease in $\mathpzc{E}_{sp}$
\begin{equation}
\label{Eq:delta_ener_EM_norm}
\frac{\Delta \mathpzc{E}_{sp}}{\mathpzc{E}_{sp}(t=0)}\sim\frac{\hat{\mathpzc{E}}_{sp}^{-}-\hat{\mathpzc{E}}_{sp}^{+}}{\hat{\mathpzc{E}}_{sp}^{+}}\sim-\frac{16c\sqrt{\epsilon}}{\omega_{pe}L}
\end{equation}
to lowest order in $\epsilon$ when transitioning from a compression to a rarefaction pulse. Here $L$ is the length of the simulation domain. For the simulation parameters given in Table~\ref{Tab:input_deck} this would yield a $5.2\%$ decrease compared to the initial pulse-slab EM energy of a compression pulse. As observed in Fig.~\ref{Fig:Energy_tot}.a, this theoretical prediction is in very good agreement with the change in EM energy observed numerically after the first reflection ($t\omega_{ci}\sim0.9$). Furthermore, simulations confirm that the energy lost during the first reflection is almost entirely gained back when transitioning back to a compression pulse after the second reflection ($t\omega_{ci}\sim2.8$). Indeed $\mathpzc{E}_{sp}$ is seen in Fig.~\ref{Fig:Energy_tot}.a to return to a value very close to its initial value. This agreement suggests that the EM energy in the system goes up and down to reflect changes in the nature of the pulse upon reflection, as illustrated schematically in Fig.~\ref{Fig:Energy}. 

This simple picture for the energy dynamics is further supported by considering the Poynting flux in Eq.~(\ref{Eq:Poynting}). To see this, let us use that $E_y^t = 2E_y^i$ and $B_z = B_0$ for a perfect reflection (\emph{i.~e.} $r_F=1$ in Eq.~(\ref{Eq:magnetic_reflection_coeff})) and introduce $\Pi_x^0 = 2{\mu_0}^{-1}E_y^iB_0$ the $x$ component of the Poynting vector constructed with these fields. Note that $\Pi_x^0$ can also be derived as the Taylor expansion of $\bm{\Pi}\cdot\mathbf{\hat{x}}$ to zero$^{th}$ order in $v_A/c$. Writing $t_1$ and $t_2$ times before and after reflection, the energy flowing away from the system through a surface of normal $\mathbf{\hat{x}}$ is accordingly
\begin{align}
\label{Eq:Poynting_zero_order}
\mathcal{E}^0 \doteq \int_{t_1}^{t_2} \Pi_x^0dt  & = \frac{B_0}{\mu_0}\int_{-\infty}^{\infty} \frac{2\epsilon v_AB_0}{\bar{\omega}}\textrm{sech}^2(\bar{t})d\bar{t}\nonumber\\
 & = 8\frac{c}{\omega_{pe}}\frac{{B_0}^2}{\mu_0}\sqrt{\epsilon},
 \end{align}
with $\bar{\omega}t = \bar{t}$ and $\bar{\omega} = \omega_{ci}\sqrt{\epsilon}/(2\eta)$. This energy $\mathcal{E}^0$ happens to be precisely the difference between the slab-pulse energy for a compression pulse $\hat{\mathpzc{E}}_{sp}^{+}$ and that for a rarefaction pulse $\hat{\mathpzc{E}}_{sp}^{-}$ as obtained in Eq.~(\ref{Eq:diff}) and rewritten as a normalized quantity in Eq.~(\ref{Eq:delta_ener_EM_norm}). Moreover, since the outward pointing normal to the closed volume surface bounding the simulation volume is $\mathbf{\hat{x}}$ on the right hand side and $-\mathbf{\hat{x}}$ on the left hand side, one recovers that the transformation of a compression pulse into a rarefaction pulse upon reflection on the right hand side is associated with an energy loss $-\mathcal{E}^0$, while the transformation of a rarefaction pulse into a compression pulse upon reflection on the left hand side is associated with an energy gain $\mathcal{E}^0$. Note that while an energy gain from a vacuum region may initially appear odd, it is important to recall that the surrounding vacuum is assumed here to be permeated by a background magnetic field $B_0\mathbf{\hat{z}}$, and therefore that the EM energy density in this region is not zero. 

Finally, the fact that the energy dynamics of the slab-pulse system is primarily associated with radiation to the vacuum region as opposed to Joule losses is confirmed by analysing separately these two contributions in numerical simulations. As shown in Fig.~\ref{Fig:Energy_tot}.b, one indeed verifies that the variation in $\mathpzc{E}_{sp}$ stems predominantly from the flux of the Poynting vector $\bm{\Pi}$ at the axial ends of the simulation domain. 

\subsection{Energy balance in the pulse alone}

Let us now analyze the dynamics of the pulse alone and consider for this the EM energy of the pulse
\begin{equation}
\mathpzc{E}_{p} \doteq \frac{1}{2}\int_V \left[\epsilon_0 |\mathbf{E}|^2+{\mu_0}^{-1}|\mathbf{B}-\mathbf{B}_0|^2\right]dV,
\end{equation}
the kinetic energy of the pulse
\begin{equation}
\mathpzc{F}_{p} \doteq \sum_{\alpha=i,e}\varepsilon_{K_{\alpha}} = \sum_{\alpha=i,e} \frac{1}{2}\int_V n_\alpha m_\alpha {v_{\alpha}}^2dV
\end{equation}
and the internal energy in the system
\begin{equation}
\varepsilon_{I} \doteq \sum_{\alpha=i,e}\varepsilon_{I_{\alpha}} = \sum_{\alpha=i,e}\frac{3}{2}\int_V n_\alpha k_B T_{\alpha}dV
\end{equation}
with $k_B$ the Boltzmann constant. The time evolution of $\mathpzc{E}_{p}$, $\mathpzc{F}_{p}$, $\varepsilon_{I}$ and the sum of these quantities is plotted in Fig.~\ref{Fig:Energy_pulse_slab}, with all quantities normalized by the energy for a soliton-like pulse $\hat{\mathpzc{E}}_{p}$ given in Tab.~\ref{Tab:soliton_var}. 

\subsubsection{Losses due to partial reflection}

\begin{figure}
\begin{center}
\includegraphics[]{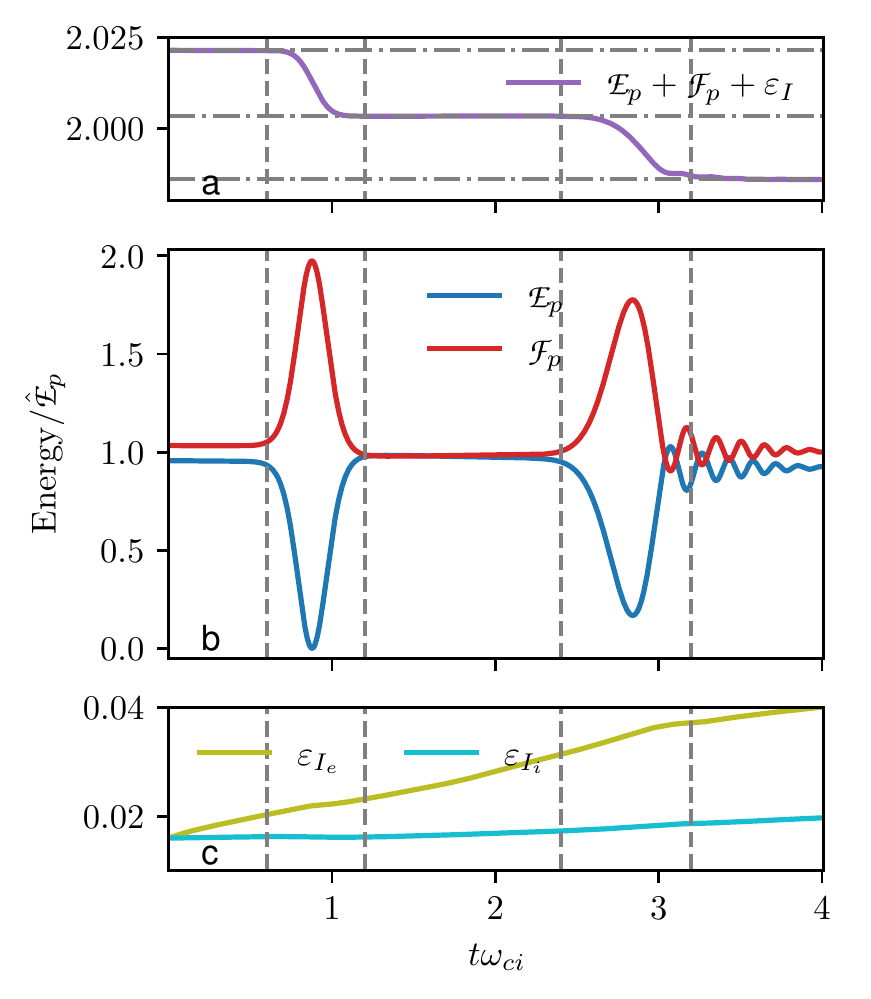}
\caption{Time evolution of the total pulse energy (a) the electromagnetic $\mathpzc{E}_{p}$ and kinetic $\mathpzc{F}_{p}$ components of the pulse energy (b) and the electron and ion internal energy $\varepsilon_{I}$ in the slab (c). Energies are normalized by the soliton EM energy $\hat{\mathpzc{E}}_p$ given in Tab.~\ref{Tab:soliton_var}. The vertical dashed gray lines depict the start and finish of reflection events, consistent with those shown in Fig.~\ref{Fig:BMap}. }
\label{Fig:Energy_pulse_slab}
\end{center}
\end{figure}

Starting from Fig.~\ref{Fig:Energy_pulse_slab}.a, one observes that the total energy $\varepsilon_{EM}+\varepsilon_K+\varepsilon_I$ is constant away from reflection events but decreases at each pulse reflection. Quantitatively, the total energy is measured to decrease by $9\permil$ and $8.6\permil$ for the first and second reflection, respectively. A point of comparison for these values can be obtained from the energy reflection coefficients $R_l = |{r_F}^2|$ and $R_s = |{r_F}^{3/2}|$ for linear and KdV soliton waves~\cite{Lonngren1991,Gueroult2018b}, respectively.  From Eq.~(\ref{Eq:magnetic_reflection_coeff}) one gets $T_l=1-R_l \sim4v_A/c\sim 9.3~10^{-3}$ and $T_s = 1-R_s = 7~10^{-3}$ for the simulation parameters given in Table~\ref{Tab:input_deck}. The very good agreement between predictions obtained from simple EM field continuity conditions at the plasma-vacuum interface and the energy loss observed in simulations already observed in Ref.~\cite{Gueroult2018b} is thus recovered here in simulations with a physical electron to ion mass ratio $\eta^2$. This is interesting since the use of a real mass ratio leads to a much smaller value of $v_A/c$ for a given overdense regime $\omega_{pe}/\omega_{ce}$, and in turn to a much weaker transmission.

The origin for this variation in total pulse energy upon reflection can be better understood by examining the flux of the Poynting vector through the simulation boundaries as done in Eq.~(\ref{Eq:Poynting_zero_order}), but by considering this time a partial reflection with $E_y^t = c/v_At_F E_y^i \sim2E_y^i(1-v_A/c)$ and $B_1^t=t_F B_1^i\sim 2v_A/cB_1^i$. Getting back to the definition of the Poynting flux in the vacuum region, one can show that
\begin{equation}
\bm{\Pi}\cdot\mathbf{\hat{x}} = \Pi_x^0+\Pi_x^{1,E}+\Pi_x^{1,B} +\mathcal{O}\left[\left(\frac{v_A}{c}\right)^2\right],
\end{equation}
with
\begin{subequations}
\begin{equation}
\Pi_x^{1,E} \doteq -\frac{v_A}{c}\Pi_x^0
\end{equation}
the first order contribution of the transmitted transverse electric field $E_y^t$ and
\begin{equation}
\Pi_x^{1,B} \doteq \pm4\frac{v_A}{c}\frac{E_y^iB_1^i}{\mu_0}
\end{equation}
\end{subequations}
the first order contribution of the transmitted magnetic field $B_1^t$. The $\pm$ sign in the definition of $\Pi_x^{1,B}$ reflects the field convention shown in Fig.~\ref{Fig:ReflectionConfig}, with $B_1$ parallel and anti-parallel to $\mathbf{\hat{z}}$ on the right- and left-hand side, respectively. 

Let us first consider the $\Pi_x^{1,E}$ contribution. This term corresponds to the correction to Eq.~(\ref{Eq:Poynting_zero_order}) associated with the fact that the rarefaction pulse no longer has an amplitude $\epsilon$. Indeed, one verifies that the lowest order expansion in $v_A/c$ of Eq.~(\ref{Eq:diff}) predicts a correction to the zero$^{th}$ order (perfect reflection) energy difference between compression and rarefaction systems $\Delta \mathpzc{E}_{sp}^0$ equal to $-v_A\Delta \mathpzc{E}_{sp}^0/c$, which is consistent with the correction of the Poynting flux $\Pi_x^{1,E}$. This implies that the energy exchange associated with to the transformation from compression to rarefaction and vice versa is $(1-v_A/c)$ smaller than $\mathcal{E}^0$.

Moving on to $\Pi_x^{1,B}$, one can show similarly to what was done in Eq.~(\ref{Eq:Poynting_zero_order}) that
\begin{align}
\label{Eq:Poynting_B_1}
\mathcal{E}^1 \doteq \int_{t_1}^{t_2} \Pi_x^{1,B}dt  & = \pm\frac{1}{\mu_0}\int_{-\infty}^{\infty} \frac{4\epsilon^2 v_A{B_0}^2}{\bar{\omega}}\frac{v_A}{c}\textrm{sech}^4(\bar{t})d\bar{t}\nonumber\\
 & = \pm\frac{32}{3}\frac{c}{\omega_{pe}}\frac{{B_0}^2}{\mu_0}\frac{v_A}{c}\epsilon^{3/2}\nonumber\\
 & = \pm4 \frac{v_A}{c}\left(\hat{\mathpzc{E}}_p+\hat{\mathpzc{F}}_p\right),
 \end{align}
where we have used the definition of the pulse EM and kinetic energy $\hat{\mathpzc{E}}_p=\hat{\mathpzc{F}}_p$ in Tab.~\ref{Tab:soliton_var}. Since, as mentioned earlier, the outward pointing normal vector changes sign when going from right to left boundary, this contribution is always positive, that is to that it is an energy loss for the slab-pulse system for all reflections as observed in simulations results. One notes that $\mathcal{E}^1$ corresponds to a relative loss of $4v_A/c$ in both mechanical and magnetic energy, which is precisely the transmitted energy predicted by Fresnel coefficients. This is remarkable in that Fresnel reflection and transmission coefficients are derived for field quantities, but it is shown here to hold equally for the mechanical energy which makes for the other half of the pulse energy in the case of a magnetosonic soliton.

Taking a step back, Eq.~(\ref{Eq:Poynting_zero_order}) and Eq.~(\ref{Eq:Poynting_B_1}) underline two different energy exchange mechanisms upon reflection, as illustrated in Fig.~\ref{Fig:Energy}. The first, associated with the Poynting flux constructed from the background magnetic field $B_0\mathbf{\hat{z}}$, corresponds the change in field energy between compression and rarefaction slab-pulse systems. It is associated with a loss (a gain) of energy by $\mathcal{E}^{0}$ for the slab-pulse system for the transformation of a compression (rarefaction) pulse into a rarefaction (compression) pulse. For a partial reflection it is corrected by a factor $1-v_A/c$. The second, associated with the Poynting flux constructed from the perturbation magnetic field $t B_1^i\mathbf{\hat{z}}$, corresponds to a loss of pulse energy (EM fields plus mechanical) by $\mathcal{E}^1$ due to a partial reflection. In this case both the EM and mechanical energy losses are found to separately follow Fresnel's transmission coefficient for the energy $1-{r_F}^2$.

\subsubsection{Field and mechanical energy partitioning}

Although the total energy simply decreases upon reflection, the examination of the time evolution of the different energy contents as plotted in Fig.~\ref{Fig:Energy_pulse_slab}.b reveals a more complex dynamics. 
Indeed, while the EM and kinetic energy are comparable away from reflections, this is clearly not the case during reflections as it can be seen in Fig.~\ref{Fig:Energy_pulse_slab}.b. As the pulse gets reflected, the EM energy first nearly goes to zero (for $t\omega_{ci}\sim 0.9$) before recovering a value close to that before reflection. This decrease and then increase in EM energy is consistent with the fact that the amplitude of the magnetic perturbation goes through zero as it changes sign. 
Now, since we have shown earlier that the total pulse energy is nearly conserved as $r_F\sim1$, this temporary decrease in EM energy $\mathpzc{E}_{p}$ must then be compensated by a growth of the kinetic energy $\mathpzc{F}_{p}$. Physically, this behaviour can be understood by recalling that the matching condition for the transverse electric field at the plasma-vacuum interface. Here, as the magnetic field perturbation is nearly zero (same amplitude but opposite sign for incident and reflected magnetic perturbations), the transverse electric field nearly doubles (same amplitude and same sign for incident and reflected transverse electric fields). This in turn leads to twice as large a longitudinal electron velocity $E_y/B_z$ and, from quasi-neutrality, twice as large an ion longitudinal velocity. This phenomenologically explains the doubling of $\mathpzc{F}_p$ observed in Fig.~\ref{Fig:Energy_pulse_slab}.b. 

For completeness, this picture where the EM energy goes momentarily to zero only applies to the first reflection. Indeed, as seen in \ref{Fig:Energy_pulse_slab}.b, the EM energy does not decrease as strongly for the second reflection ($t\omega_{ci}\sim 2.8$). This is because the magnetic perturbation after the first reflection is no longer a single pulse but instead a pulse followed by a trailing wave, as shown in Fig.~\ref{Fig:BMap}. We note though that even in this case the momentarily decrease in EM energy $\mathpzc{E}_{p}$ is fully compensated, other than for the losses due to $r$ being slightly smaller than $1$, by an increase in kinetic energy $\mathpzc{F}_{p}$.

\subsubsection{Dissipation}

\begin{figure}
\begin{center}
\includegraphics[]{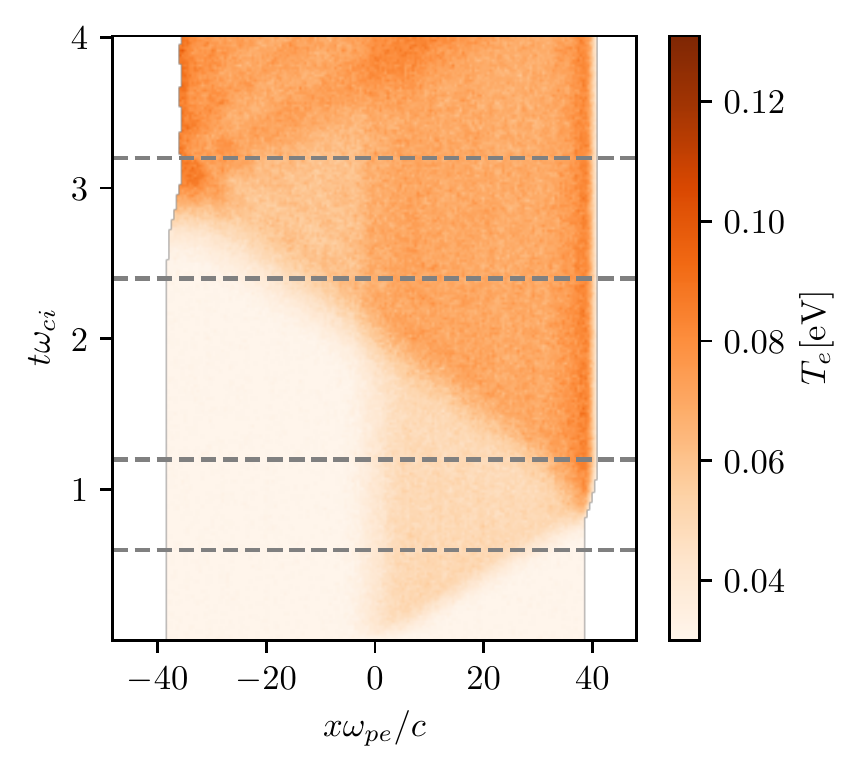}
\caption{Map of the electron temperature. A clear increase of temperature is observed in the wake of the pulse. The horizontal dashed lines approximately indicate the start and finish of reflection events. }
\label{Fig:Temap}
\end{center}
\end{figure}

Away from reflection events, a closer examination shows that the EM energy slowly decreases as the pulse propagates through the plasma slab. Meanwhile, the electron internal energy $\epsilon_{I_e}$ grows, as shown in Fig.~\ref{Fig:Energy_pulse_slab}.c. This increase in electron internal energy reflects an increase in $T_e$ in the wake of the pulse as shown in Fig.~\ref{Fig:Temap}. We interpret this as the signature of a weak energy transfer between the pulse and the plasma slab, even away from reflections. Also this behaviour was already noticed in our earlier simulations~\cite{Gueroult2018b}, the artificial mass ratio used in this case did not make it possible to conclude on the deposition mechanism. The present simulations allow postulating the following scenario. 

Consider the effect of electron-ion Coulomb collisions on the soliton pulse and write $\nu_{ei}$ the electron-ion momentum transfer frequency. The local governing equation for the electron temperature writes
\begin{equation}
\frac{3}{2}k_B\frac{d}{dt}(n_eT_e) = \frac{m_e\nu_{ei}}{n_ee^2}{|\bm{j}|}^2-3\eta^2n_e\nu_{ei}(T_e-T_i).
\label{Eq:electron_temperature}
\end{equation}
The first term on the right hand side stems from the friction force on electrons flowing relative to ions. For the soliton structure governed by Eqs.~(\ref{Eq:Soliton_scaling}), $v_{e_x}\sim v_{i_x}$ but $v_{e_y}\neq v_{i_y}$ so that the friction force is along $\mathbf{\hat{y}}$ and $|\bm{j}|^2\sim {j_y}^2\sim (n_e e v_{e_y})^2$. The second term on the right hand side in Eq.~(\ref{Eq:electron_temperature}) is the energy transfer from electrons to ions.

While energy transfer from electrons to ions was observed to play a role in earlier simulations~\cite{Gueroult2018b}, it is expected to be less important here due to the use of a physical mass ratio $\eta^2$. This hypothesis is confirmed by the weak variation of the ion internal energy shown in Fig.~\ref{Fig:Energy_pulse_slab}.c. In addition, since $j_y$ is zero outside of the pulse, a negligible energy transfer between electron and ion should lead to no significant change in $T_e$ away from the pulse. This is indeed what is observed in Fig.~\ref{Fig:Temap}, supporting further that the dominant term on the right-hand side of Eq.~(\ref{Eq:electron_temperature}) is Joule heating ${j_y}^2/\sigma$ with $\sigma = n_e e^2/(m_e\nu_{ei})$ the conductivity. Integrating the Joule heating over the low amplitude pulse profile travelling at velocity $v_A$ Eqs.~(\ref{Eq:Soliton_scaling}) then leads to a variation in temperature after the pulse passage
\begin{equation}
\frac{3}{2}k_B\Delta T_e = \frac{8}{15}\frac{m_iv_A\nu_{ei}c}{\omega_{pe}}\epsilon^{5/2}
\end{equation}
or, equivalently, a variation in the system's electron internal energy 
\begin{align}
\Delta \varepsilon_{I_e} \doteq & \varepsilon_{I_e}(t) - \varepsilon_{I_e}(t=0) \nonumber\\
= & \frac{8}{15}\frac{n_0m_i{v_A}^2c}{\omega_{pe}}\nu_{ei}\epsilon^{5/2}t.
\label{Eq:analytical_internal_energy_variation}
\end{align}  

\begin{figure}
\begin{center}
\includegraphics[]{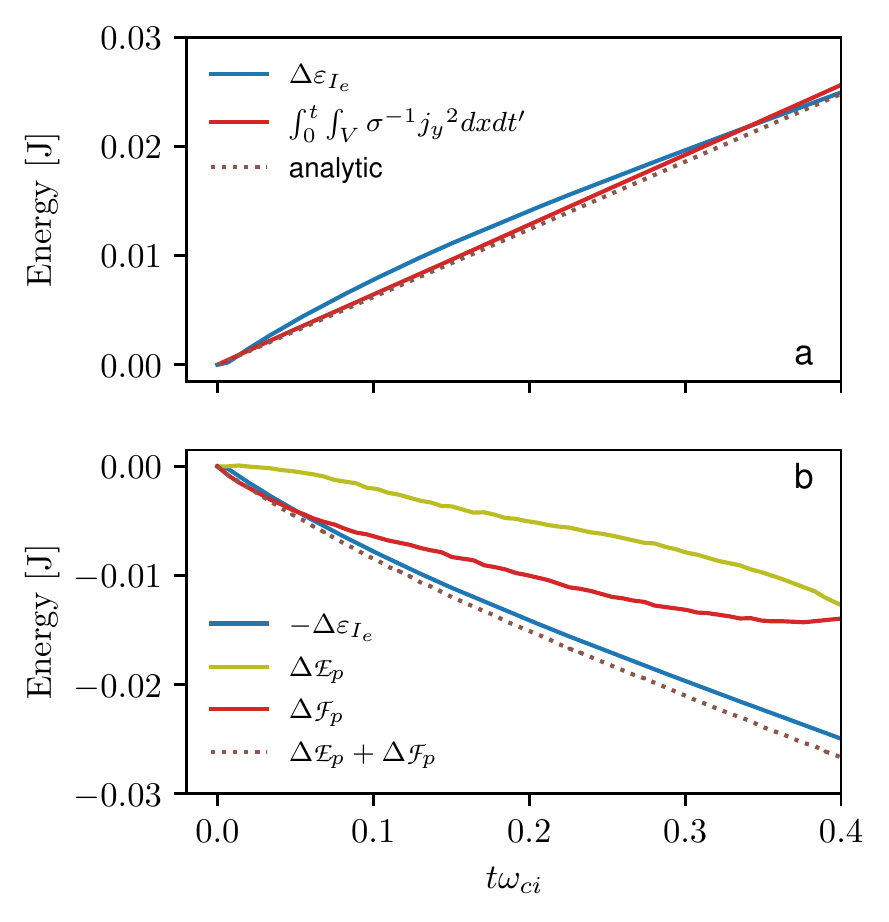}
\caption{Time evolution of the pulse energy content before the first reflection event.  (a) Electron heating $\Delta \varepsilon_{I_e}$ follows the non-zero ${j_y}^2/\sigma$ volumic resistive power losses in the pulse, as expected from Eq.~(\ref{Eq:electron_temperature}), and both quantities can be reasonably matched using an electron-ion momentum transfer frequency $\nu_{ei}^{\star}=2.6~10^7$~s$^{-1}$. The analytic solution corresponds to Eq.~(\ref{Eq:analytical_internal_energy_variation}) (b) The corresponding pulse energy loss $-\Delta \varepsilon_{I_e}$ is observed to be evenly split between the mechanical and electromagnetic energy contents $\mathpzc{F}_{p}$ and $\mathpzc{E}_{p}$, respectively.   }
\label{Fig:JouleHeating}
\end{center}
\end{figure}

Evidence supporting this scenario is plotted in Fig.~\ref{Fig:JouleHeating}.a. It is indeed found that the variation in electron internal energy $\Delta \varepsilon_{I_e}$ can be appropriately matched to both the time and volume integrated power density ${j_y}^2/\sigma$ and the analytical formula given in Eq.~(\ref{Eq:analytical_internal_energy_variation}) using a constant electron-ion momentum transfer frequency $\nu_{ei}$. The best fit shown in Fig.~\ref{Fig:JouleHeating}.a is obtained for $\nu_{ei}^{\star}=2.6~10^7$~s$^{-1}$. Analysis of the time evolution of the different contributions to the pulse energy further shows, as plotted in Fig.~\ref{Fig:JouleHeating}.b, that the energy spent heating electrons is split nearly evenly between the mechanical pulse energy $\mathpzc{F}_{p}$ and electromagnetic pulse energy $\mathpzc{E}_{p}$. This last result is consistent with the fact that for a soliton $\hat{\mathpzc{E}}_{p} = \hat{\mathpzc{F}}_{p}$ to lowest order in $\epsilon$. From Eqs.~(\ref{Eq:EM_energy_pulse}), (\ref{Eq:K_energy_pulse}) and (\ref{Eq:analytical_internal_energy_variation}), one then finds that 
\begin{equation}
\frac{1}{\mathpzc{F}_{p}^{\diamond}}\frac{d\mathpzc{F}_{p}}{dt} = \frac{1}{\mathpzc{E}_{p}^{\diamond}}\frac{d\mathpzc{E}_{p}}{dt} = -\frac{\epsilon\nu_{ei}^{\star}}{5},
\end{equation}
where $u^{\diamond} = u(t=0)$.

\begin{figure}
\begin{center}
\includegraphics[]{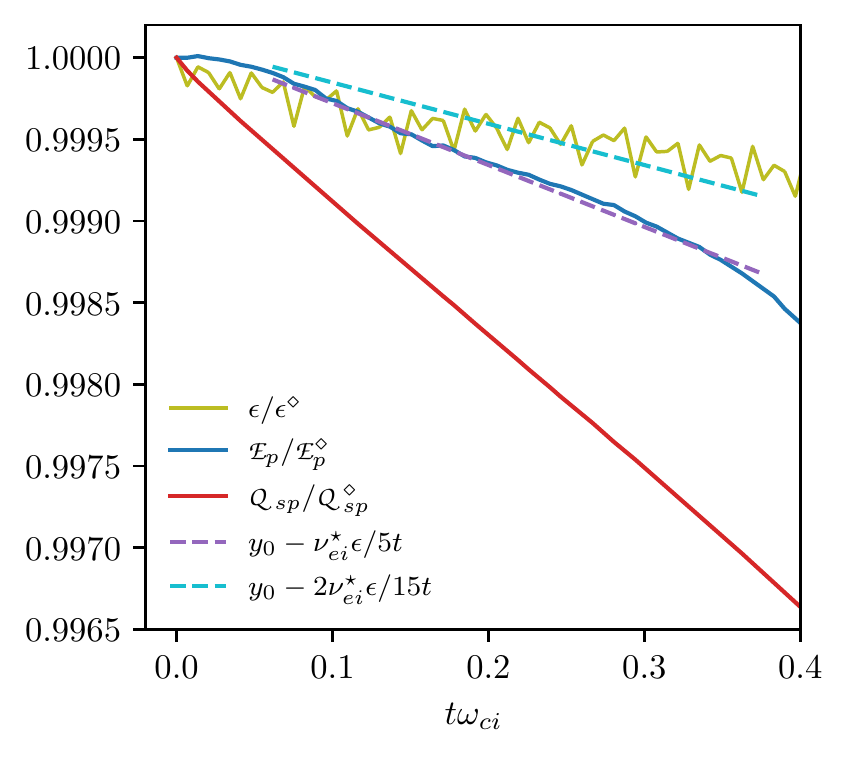}
\caption{Time evolution of normalized pulse field energy $\mathpzc{E}_p$, pulse amplitude $\epsilon$ and pulse mechanical momentum $\mathpzc{Q}\,_{sp}$ before the first reflection event. The decrease rate for $\mathpzc{E}_p$ and $B_1$ are in relatively good agreement with the values derived from electron-ion friction and the soliton scaling (dashed curves). On the contrary, $\mathpzc{Q}\,_{sp}$ decreases much faster than the $-
\nu_{ei}^{\star}\epsilon/15t$ obtained based on the pulse amplitude decrease.  }
\label{Fig:LossScaling}
\end{center}
\end{figure}

Exploring further the possible implications of Coulomb collisions and recalling that $\mathpzc{E}_{p}\propto \epsilon^{3/2}$, a pulse satisfying to the soliton scaling should in principle see a normalized decrease in amplitude due to collisions
\begin{equation}
\frac{d}{dt}\left(\frac{\epsilon}{\epsilon^{\diamond}}\right)  = -\frac{2}{3}\sqrt{\frac{\epsilon^{\diamond}}{\epsilon}} \frac{1}{\mathpzc{E}_{p}^{\diamond}}\frac{d\mathpzc{E}_{p}}{dt}.
\end{equation}
Similarly, since $\hat{\mathpzc{Q}\,}^{\pm}\propto \epsilon^{1/2}$, this decrease in pulse amplitude should lead to a normalized loss of mechanical longitudinal momentum  
\begin{equation}
\frac{1}{\mathpzc{Q}\,_{sp}^{\diamond}}\frac{d\mathpzc{Q}\,_{sp}}{dt} = -\frac{1}{3}\sqrt{\frac{\epsilon}{\epsilon^{\diamond}}} \frac{1}{\mathpzc{E}_{p}^{\diamond}}\frac{d\mathpzc{E}_{p}}{dt},
\label{Eq:momentum_decrease_rate}
\end{equation}
where we have defined the mechanical longitudinal momentum in the simulation
\begin{equation}
\label{Eq:mech_mom_def}
\mathpzc{Q}\,_{sp} \doteq \sum_{\alpha=i,e}\int_V n_\alpha m_\alpha {v_{\alpha_x}}dV.
\end{equation}
As shown in Fig.~\ref{Fig:LossScaling}, the decrease in pulse magnetic energy and in pulse amplitude roughly follow these scalings, supporting the role of collisions. On the other hand, the decrease in $\mathpzc{Q}\,_{sp}$ which is predicted from Eq.~(\ref{Eq:momentum_decrease_rate}) to be slower than that of the pulse field energy $\mathpzc{E}_p$, is observed to be much faster in simulations. This possibly points to an additional contribution to the momentum dynamics, and this possibility is analysed in more details in Sec.~\ref{Sec:Momentum}.

\subsubsection{Collisionality}

An important remark that is called for at this point is that PIC simulations presented in this study are done without modelling Coulomb collisions. The collisional signatures observed here are a purely numerical effect stemming from the finite size particles in PIC models~\cite{Okuda1970}. In actuality, the electron-ion momentum transfer collision rate 
\begin{equation}
\nu_{ei} = \frac{n_0e^4{Z_i}^2\ln{\Lambda}}{4\pi{\epsilon_0}^2\sqrt{m_e}{T_e}^{3/2}}
\end{equation}
computed for the plasma parameters given in Table~\ref{Tab:input_deck} is orders of magnitude larger than $\nu_{ei}^{\star}$ (between $10^6$ and $10^4$ larger depending on whether one uses the transverse electron velocity or the electron thermal speed). One could then argue that the results discussed here are non-physical since a much higher collision rate would significantly augment the effects observed here. 

Although this is true strictly speaking, we stress again that our goal is to examine the physics of over-dense plasmas $\omega_{pe}/\omega_{ce}\gg1$, but that the choice of plasma parameters used here and in particular of plasma density is only dictated by PIC requirements. For the same $\omega_{pe}/\omega_{ce}$ but a lower plasma density, the effect of collisions will be comparatively weaker, and negligible for a sufficiently small density. This is particularly true when considering that the initial temperatures $T_e = T_i$ in the simulation are primarily dictated by the requirement that $v_{th_i}$ is small compared to typical ion velocities in the pulse. Since these velocities scale with $v_A\propto n^{-1/2}$, a decrease in density allows for higher temperatures. Quantitatively, a decrease in density by a factor $\varsigma$ leads to an increase in $v_A$ by a factor $\varsigma^{1/2}$. This allows for an increase in temperature by a factor $\varsigma$ and, as a a result, in a decrease in $\nu_{ei}$ by a factor $\varsigma^{5/2}$. Meanwhile, the pulse transit time across the domain remains unchanged since it scale as $c/(\omega_{pe}v_A)$. This suggests the $1\%$ decrease in pulse energy over one pulse transit observed in simulations could be obtained by taking $\varsigma \sim 100$ while keeping $\omega_{pe}/\omega_{ce}$ constant (\emph{i.~e.} decreasing $B_0$ by a factor $\varsigma^{1/2}\sim10$). Taking $\varsigma \sim 1000$ would yield a decrease in pulse energy due to Coulomb collisions of less than $1\permil$. In other words, the physics studied here is expected to be representative of that of over-dense regimes in the limit that plasma densities are low enough for collisional effects to be negligible. 

\section{Momentum}
\label{Sec:Momentum}

\subsection{Mechanical and EM momentum dynamics}

\subsubsection{Definitions}

Let us recall here the definition of the mechanical momentum
\begin{equation*}
\mathpzc{Q}\,_{sp} \doteq  \sum_{\alpha=i,e}\int_V n_\alpha m_\alpha {v_{\alpha_x}}dV.
\end{equation*}
given in Eq.~(\ref{Eq:mech_mom_def}). We similarly define the pulse momentum along $\mathbf{\hat{x}}$
\begin{equation}
\mathpzc{P}_p \doteq \epsilon_0  \int_V  E_y B_1dV 
\end{equation}
and the slab-pulse EM momentum along $\mathbf{\hat{x}}$
\begin{equation}
\mathpzc{P}_{sp} \doteq \epsilon_0\int_V E_y (B_0+B_1)dV,
\end{equation}
with $V$ a volume just enclosing the plasma slab. For a soliton-like pulse in an infinite plasma these momenta are respectively equal to $\hat{\mathpzc{Q}\,}$, $\hat{\mathpzc{P}}_p^{\pm}$ and $\hat{\mathpzc{P}}_{sp}$, as given in Tab.~\ref{Tab:soliton_var}.

\subsubsection{Observations}

\begin{figure}
\begin{center}
\includegraphics[]{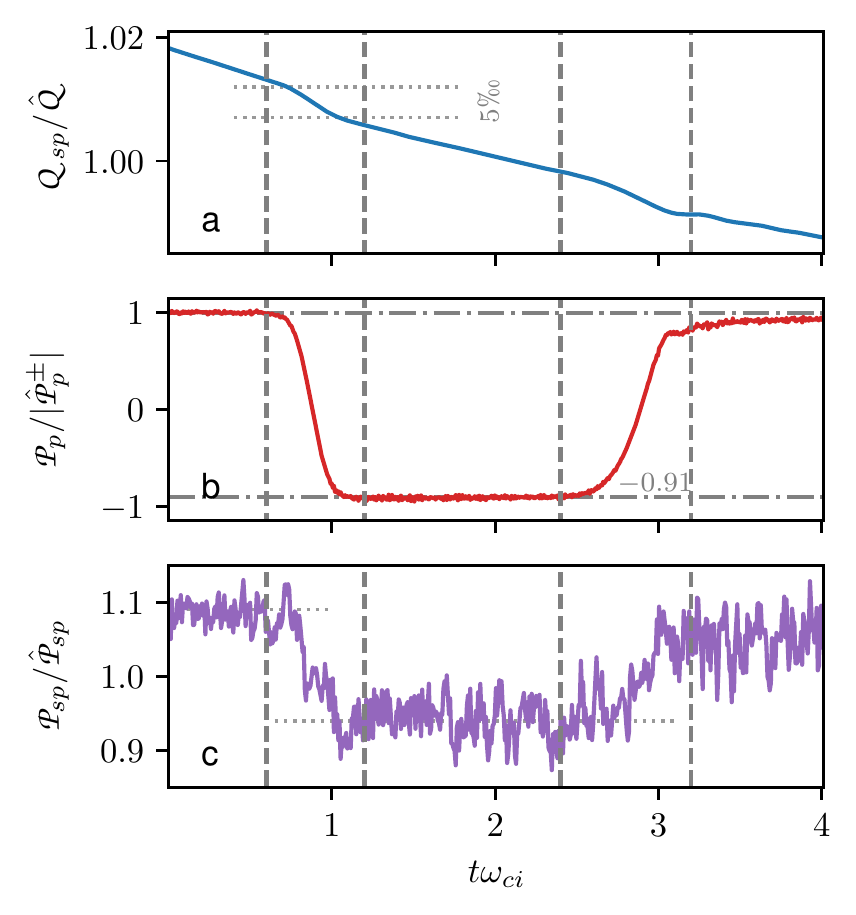}
\caption{Time evolution of the mechanical momentum $\mathpzc{Q}\,_{sp}$ (a) and pulse $\mathpzc{P}_p$ (b) and slab-pulse $\mathpzc{P}_{sp}$ (c) electromagnetic momentum. Momenta are normalized by their soliton scaling $\hat{\mathpzc{Q}\,}$, $|\hat{\mathpzc{P}}_p^{\pm}|$ and $\hat{\mathpzc{P}}_{sp}$ given in Tab.~\ref{Tab:soliton_var}. The vertical dashed gray lines depict the start and finish of reflection events, consistent with those shown in Fig.~\ref{Fig:BMap}. }
\label{Fig:Momentum}
\end{center}
\end{figure}

The time evolution of these three momenta plotted in Fig.~\ref{Fig:Momentum} immediately confirms some basic qualitative properties of reflection that were anticipated from the soliton picture derived in Sec.~\ref{Sec:Soliton_scalings}. 

First, the pulse EM longitudinal momentum $\mathpzc{P}_p$ changes sign as the pulse is reflected (Fig.~\ref{Fig:Momentum}.b), whereas both the mechanical longitudinal momentum $\mathpzc{Q}\,_{sp}$ and the pulse-slab EM longitudinal momentum $\mathpzc{P}_{sp}$ (Fig.~\ref{Fig:Momentum}.a and Fig.~\ref{Fig:Momentum}.c, respectively) remain positive. This is consistent with the analytical results that $|\hat{\mathpzc{P}}_p^{\pm}|$ depends on the nature of the soliton, whereas $\hat{\mathpzc{Q}\,}$ and $\hat{\mathpzc{P}}_{sp}$ do not to lowest order in $\epsilon$. 

Looking more closely at the mechanical momentum, Fig.~\ref{Fig:Momentum}.a shows that $\mathpzc{Q}\,_{sp}$ decreases linearly with time, with an enhanced loss rate during reflection events. A loss of mechanical momentum upon reflection is qualitatively consistent with the loss in pulse amplitude expected from a partial reflection. As shown in Fig.~\ref{Fig:Momentum}.a., reflection is here observed to lead to a relative momentum loss of about $5\permil$. In contrast, the observation of a mechanical momentum loss away from reflection begs questions. Although the slow decrease in pulse amplitude due to friction can explain part of this result, it has been shown in Sec.~\ref{Sec:Energy} than it can not yield the observed decrease rate alone. Looking for possible explanations, we show in Appendix~\ref{Sec:App1} that this rate is remarkably consistent with the negative longitudinal momentum that an ion in the wake of the travelling pulse will exhibit on timescales short compared to the ion gyro-period. However, this model of higher order corrections to the transverse ion dynamics in a soliton pulse can not explain that the loss of momentum is linear with time. 
A complete picture of the mechanism behind this momentum loss away from reflections, and especially ahead of the first reflection while the pulse is close to a soliton, is thus still lacking. 

Finally, examining now the pulse and pulse-slab EM momenta in Fig.~\ref{Fig:Momentum}.b and Fig.~\ref{Fig:Momentum}.c, we find that in contrast with mechanical momentum these two quantities remain nearly constant in between reflection events. Looking at reflections, momentum is lost when going from a compression to a rarefaction pulse (first reflection) and gained when going from a rarefaction to a compression pulse (second reflection). Quantitatively, the loss in EM momentum measured in simulations after first reflection is about $9\%$ and $13\%$ for $|\mathpzc{P}_p|$ and $\mathpzc{P}_{sp}$, respectively, which is again surprisingly consistent with soliton scaling. Indeed, since the pulse momentum is in the $-\mathbf{\hat{x}}$ direction after reflection, the relative change in $\mathpzc{P}_p$ is then about $-1.91$, which, from the scaling $\hat{\mathpzc{P}}_{sp} = 3\epsilon^{-1}|\hat{\mathpzc{P}}_p^{\pm}|/2$ given in Eq.~(\ref{Eq:momentum_ratio}), predicts a relative loss in pulse-slab momentum of $12.7\%$. The change in pulse-slab EM momentum thus directly reflects the change in pulse EM momentum. This last result is consistent with the fact that the next order in $\epsilon$ correction to $\hat{\mathpzc{P}}_{sp}$ in Eq.~(\ref{Eq:momentum_EM_tot}) involves $\hat{\mathpzc{P}}_p^{\pm}$.

\subsection{Momentum fluxes partitioning}

To better understand the momentum dynamics observed in simulations and described in the previous paragraph, we now look at the momentum balance in the system using the soliton results obtained in Sec.~\ref{Sec:Soliton_scalings}.  

\subsubsection{Momentum fluxes}

The local conservation equation for the total momentum along $\mathbf{\hat{x}}$ writes
\begin{align}
\frac{\partial}{\partial t}\left(n_pm_p v_{i_x}\right. + & \left.\epsilon_0 E_y B_z\right) = -\frac{\partial}{\partial x}\left(n_pm_p {v_{i_x}}^2+p\right)\nonumber\\
 &-\frac{1}{2}\frac{\partial}{\partial x}\left(\epsilon_0\left[{E_y}^2-{E_x}^2\right]+\frac{{B_z}^2}{\mu_0}\right),
\label{Eq:momentum_conservation_maxwell}
\end{align}
with $p$ the pressure. In writing Eq.~(\ref{Eq:momentum_conservation_maxwell}) we have used that since $\partial(\cdot)/\partial y =\partial(\cdot)/\partial z=0$ 
\begin{equation}
(\bm{\nabla}\cdot\mathbf{T})_x = \frac{\partial}{\partial x}T_{xx},
\end{equation}
with $\mathbf{T}$ the usual Maxwell stress tensor such that
\begin{equation}
\label{Eq:Maxwell_stress_tensor_def}
T_{ij} = \epsilon_0\left[E_iE_j-\frac{1}{2}E^2\delta_{ij}\right]+{\mu_0}^{-1}\left[B_iB_j-\frac{1}{2}B^2\delta_{ij}\right].
\end{equation}
Integrating over a volume $V$ just enclosing the plasma slab $-a< x< b$ and taking advantage that on the closed surface $\partial V$ bounding this volume the density is zero and $E_x=0$, Eq.~(\ref{Eq:momentum_conservation_maxwell}) yields
\begin{align}
\label{Eq:mom_conservation_all}
\frac{d}{dt}\left[\mathpzc{Q}\,_{sp}+\mathpzc{P}_{sp}\right] & = \mathcal{T}\left(-a,t\right)-\mathcal{T}\left(b,t\right)\nonumber\\
& = D_f (\mathcal{T})
\end{align} 
where we write $\mathcal{T}= \mathcal{T}_E+\mathcal{T}_B$ with
\begin{subequations}
\begin{align}
\mathcal{T}_E & \doteq \frac{1}{2}\epsilon_0 {E_y}^2,\\
\mathcal{T}_B & \doteq \frac{{B_1}^2}{2\mu_0}+\frac{B_0B_1}{\mu_0}.
\end{align} 
\end{subequations}
Note that since Eq.~(\ref{Eq:mom_conservation_all}) takes the difference $D_f(\mathcal{T})$ of the surface integrated momentum flux $\mathcal{T}$ at the two ends of the integration volume, we purposely choose not to include the constant term $(B_0)^2/(2\mu_0)$ from Eq.~(\ref{Eq:Maxwell_stress_tensor_def}) in our definition of $\mathcal{T}_B$ as it cancels out. 

\subsubsection{Momentum exchange for perfect reflection}

Let us first examine the case of perfect reflection, that is $E_y^t = 2 E_y^i$ and $B_1^t= 0$. In this limit, which corresponds to the zero$^{th}$ order in $v_A/c$ Taylor expansion of the momentum fluxes, one immediately gets $\mathcal{T}_B^0=0$. Any change in momentum will thus result from the surface integrated momentum flux associated $\mathcal{T}_E^0=2\epsilon_0 {E_y^i}^2$. Writing $t_1$ and $t_2$ times before and after reflection, the momentum lost during reflection is 
\begin{align}
\mathcal{P}^0 & \doteq \int_{t_1}^{t_2} \mathcal{T}_E^0dt  = \frac{\epsilon_0}{2} \int_{-\infty}^{\infty} \frac{(2\epsilon v_AB_0)^2}{\bar{\omega}}\textrm{sech}^4(\bar{t})d\bar{t}\nonumber\\
 & = \frac{16}{3}\frac{\epsilon^{3/2}v_A{B_0}^2}{\mu_0\omega_{pe}c}\nonumber\\
 & = 2 |\hat{\mathpzc{P}}_p^{\pm}|.
\end{align}
Since the outward normal of the simulation is $\mathbf{\hat{x}}$ and $-\mathbf{\hat{x}}$ on the right- and left-hand side of the domain, respectively, this is a loss of momentum on the right hand side but a gain on the left hand side. This result corresponds exactly to the perfect reflection of the pulse with EM longitudinal momentum $|\hat{\mathpzc{P}}_p^{\pm}|$, and matches qualitatively the behaviour observed in Fig.~\ref{Fig:Momentum}.b and Fig.~\ref{Fig:Momentum}.c.

\subsubsection{Losses due to partial reflection}

Let us now examine, as we have done earlier for the energy, the effect of a partial reflection. Taylor expanding the surface integrated momentum fluxes, one gets $\mathcal{T}_E = \mathcal{T}_E^0 +\mathcal{T}_E^1+\mathcal{O}[(v_A/c)^2]$ and $\mathcal{T}_B = \mathcal{T}_B^1+\mathcal{O}[(v_A/c)^2]$
with 
\begin{subequations}
\begin{equation}
\mathcal{T}_E^1 \doteq - 2\frac{v_A}{c}\mathcal{T}_E^0
\end{equation}
and
\begin{equation}
\mathcal{T}_B^1 \doteq 2\frac{v_A}{c}\frac{B_0 B_1^i}{\mu_0}.
\end{equation}
\end{subequations}
From Eq.~(\ref{Eq:mom_conservation_all}), the variations in total longitudinal momentum of the slab-pulse system over a reflection that are associated with these surface integrated momentum flux contributions are 
\begin{subequations}
\begin{align}
\mathcal{P}^{1,E} & \doteq \int_{t_1}^{t_2} \mathcal{T}_E^1dt = -2\frac{v_A}{c}\mathpzc{P}^{0}\nonumber\\
 & = -4\frac{v_A}{c} |\hat{\mathpzc{P}}_p^{\pm}|
\end{align}
and
\begin{align}
\label{Eq:P1B}
\mathcal{P}^{1,B} & \doteq \int_{t_1}^{t_2} \mathcal{T}_B^1dt   = 2\frac{B_0}{\mu_0}\frac{v_A}{c} \int_{-\infty}^{\infty} \frac{\epsilon B_0}{\bar{\omega}}\textrm{sech}^2(\bar{t})d\bar{t} \nonumber\\
 & = 8 \frac{\eta{v_A}{B_0}^2\sqrt{\epsilon}}{\mu_0c\omega_{ci}}\nonumber\\
  & = 2\frac{v_A}{c} \hat{\mathpzc{Q}\,}.
\end{align}
\end{subequations}
Because the direction of the bounding surface outward normal is flipped when going from left to right boundaries, $\mathcal{P}^{1,E}$ is a momentum gain on the right hand side but a momentum loss on the left hand side. On the other hand $\mathcal{P}^{1,B}$ is always a momentum loss as the sign of the magnetic perturbation cancels out with the one of the surface normal.

Examining first $\mathcal{P}^{1,B}$, one finds that it happens to match the transmitted momentum constructed naively from the incident mechanical momentum $\hat{\mathpzc{Q}\,}$ and the Fresnel's magnetic field transmission coefficient $t_F=2 \kappa^{1/2}/(\kappa^{1/2}+1)\sim2v_A/c$. But since, as shown in Eq.~(\ref{Eq:mecha_momentum_density}), the mechanical momentum density is proportional to the pulse amplitude, $\mathcal{P}^{1,B}$ also matches the hypothetical transmitted mechanical momentum predicted based on Fresnel's transmission coefficient, and that even if the mechanical momentum density is clearly zero in the vacuum region. Very interestingly, $2v_A/c\sim4.7~10^{-3}$ for the simulation parameters, which is in remarkably good agreement with the decrease of $\mathpzc{Q}\,_{p}$ by about $5\permil$ observed upon reflection in Fig.~\ref{Fig:Momentum}.a. 

\begin{figure}
\begin{center}
\includegraphics[width=8.6cm]{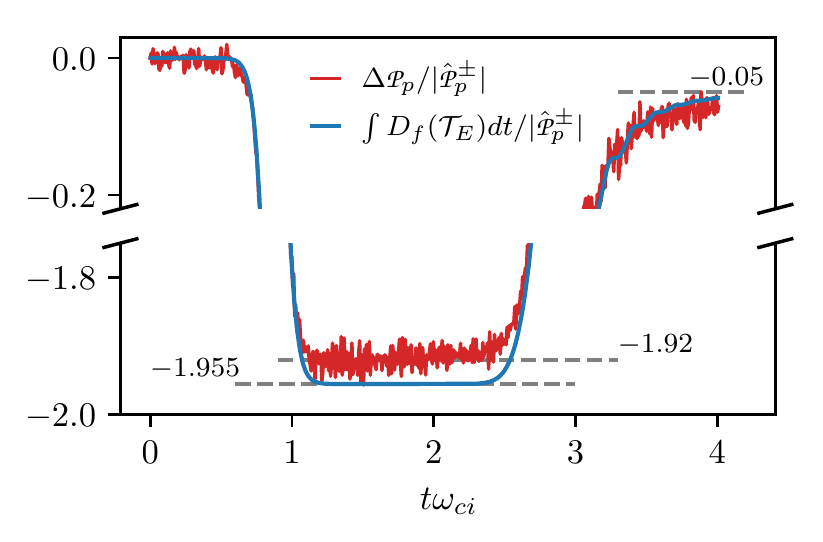}
\caption{Time integrated EM longitudinal momentum flux through the boundaries $\int D_f(\mathcal{T}_E) dt$ (electric contribution only, see Eq.~(\ref{Eq:mom_conservation_all})) and variation in pulse EM longitudinal momentum $\mathpzc{P}_p$. The amplitude of the momentum loss (gain) through the simulation boundaries upon reflection is greater than the pulse EM longitudinal momentum loss (gain) upon first (second) reflection, suggesting a momentum exchange with the plasma. Quantities are normalized by the pulse EM momentum $|\hat{\mathpzc{P}}_p^{\pm}|$ from Tab.~\ref{Tab:soliton_var}. The $y$-axis is cut to highlight the dynamics and differences between predictions and numerical results away from reflection events. }
\label{Fig:Fluxes}
\end{center}
\end{figure}

Looking now at $\mathcal{P}^{1,E}$, this term can be interpreted as a correction to $\mathcal{P}^{0}$ associated with the fact that, for a partial reflection, the amplitude of the EM longitudinal momentum of the reflected pulse $|\mathpzc{p}^r|$ is lower than that of the incident pulse $|\mathpzc{p}^i|$. Quantitatively, one gets $4v_A/c\sim9.3~10^{-3}$ for the parameters used in the simulations (see Table~\ref{Tab:dimensionless}), which suggests a reflected momentum smaller than that of the perfect reflection by less than $1\%$. Looking at simulations results, one indeed finds a decrease in reflected momentum, but the change in pulse EM momentum by $-1.92|\mathpzc{P}_p^{\pm}|$ observed in Fig.~\ref{Fig:Momentum}.b yields a variation of about $8\%$. Part of the reason for this discrepancy is likely to be found in the very idealistic nature of the soliton reflection model used here. This is supported by the finding that in simulations, as shown in Fig.~\ref{Fig:Fluxes}, the flux $\mathcal{T}_E$ through the right hand side boundary integrated over the duration of the first reflection is about $-1.955|\hat{\mathpzc{P}}_p^{\pm}|$, which is larger than the predicted $-2(1-2v_A/c)\hat{\mathpzc{P}}_p^{\pm}$. But this last result and the fact that
\begin{equation}
\mathpzc{p}^i < \mathpzc{p}^r+\mathcal{P}^{0}+\mathcal{P}^{1,E}
\end{equation}
also suggests that some EM momentum is transferred to (or from) the plasma during reflection.    
Such a transfer is indeed expected to arise from a discontinuity of $T_{xx}$ through an interface~\cite{Bisognano2017,Saldanha2010}, and while we consider the case of normal incidence for which $E_y$ and $B_z$ are continuous through the interface, the pulse exhibits a longitudinal field $E_x$ which is not continuous through the interface. Since the signs in front of ${E_x}^2$ and ${E_y}^2$ in Eq.~(\ref{Eq:momentum_conservation_maxwell}) differ, this additional mechanism should yield to a gain in pulse EM longitudinal momentum when EM momentum is lost to the vacuum upon reflection, and a momentum loss when EM momentum is gained from the vacuum upon reflection. This is indeed what is observed in Fig.~\ref{Fig:Fluxes}. However, the fact that the mechanical momentum is $(c/v_A)^2\sim10^5$ larger than the EM momentum prohibits searching for signs of this momentum transfer in simulations.

\subsubsection{Momentum balance}

\begin{figure}
\begin{center}
\includegraphics[]{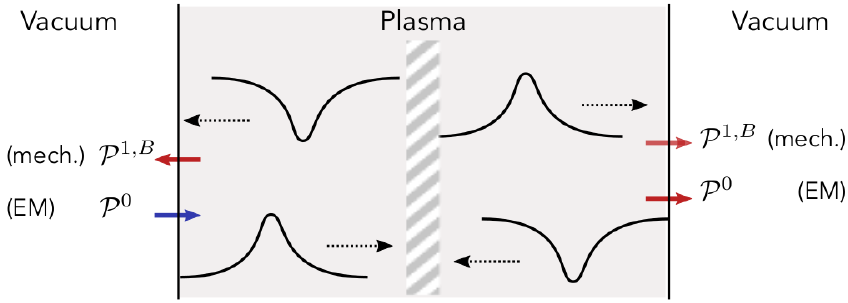}
\caption{Momentum exchange upon transformation from a rarefaction (compression) to a compression (rarefaction) pulse at the plasma-vacuum interface. For the slab-pulse system, momentum fluxes to the surrounding vacuum region can be traced back to a loss by $\mathcal{P}^{1,B}$ in mechanical momentum $\mathpzc{Q}\,_{sp}$ at each partial reflection, and a loss (a gain) by $\mathcal{P}^{0}$ in EM momentum $\mathpzc{P}_p$ when reflecting a compression (rarefaction) pulse into a rarefaction (compression) pulse.}
\label{Fig:MomentumSketch}
\end{center}
\end{figure}

Putting these pieces together, we conclude, as illustrated in Fig.~\ref{Fig:MomentumSketch}, that the mechanical momentum loss observed at each partial reflection corresponds to a transfer of positive longitudinal momentum $\mathcal{P}^{1,B}$ to the surrounding vacuum.  Supplementing this mechanical momentum loss, there is a weaker momentum transfer associated with the EM momentum of the slab-pulse system. A momentum $\mathcal{P}^{0}+\mathcal{P}^{1,E}$ is lost during the reflection of the compression pulse on the right boundary but gained upon reflection of the rarefaction pulse at the left boundary, which happens to roughly match the change in pulse EM momentum due to the change in pulse propagation direction. 

Since $\hat{\mathpzc{Q}\,}\gg\hat{\mathpzc{P}}_{sp}$, we find that the momentum loss at each partial reflection matches with good accuracy the momentum loss predicted from Fresnel's transmission coefficient $t_F$ in Eq.~(\ref{Eq:magnetic_transmission_coeff}) and the total momentum in the slab-pulse system. Recalling that the total momentum matches Minkowski's momentum $\mathcal{G}_M$, the momentum loss at each partial reflection appears to be well approximated by $t_F\mathcal{G}_M$.

Finally, we observe in simulations that the variation in mechanical and EM momentum of the system upon reflection $\Delta\mathpzc{Q}\,_{sp}$ and $\Delta\mathpzc{P}\,_{sp}$ are in good agreement with the time integral of respectively the magnetic $\mathcal{T}_B$ and electric $\mathcal{T}_E$ parts of the momentum flux density, that is
\begin{equation}
\label{Eq:momentum_partitioning}
\Delta\mathpzc{Q}\,_{sp} \sim \int_{t_1}^{t_2}\mathcal{T}_Bdt \quad\textrm{and}\quad \Delta\mathpzc{P}_{sp}\sim \int_{t_1}^{t_2}\mathcal{T}_Edt.
\end{equation} 
Interestingly, we note that this result matches the prediction obtained when considering the interface between two homogeneous isotropic dielectrics (see Appendix~\ref{Sec:App5}) in the limit of a large relative permittivity $\epsilon_R= 1+\chi$. Examining in detail the possibility of a broader validity of this prediction for the magnetized plasma configuration studied is however beyond our capabilities since it would require determining $\Delta\mathpzc{Q}\,_{sp}$ to within a factor ${\epsilon_{\perp}}^{-1}$, which is not realistic for the large value of $\chi_{\perp}\sim (c/v_A)^2$ used in simulations here.

\subsection{Insights into momentum partitioning}

Although we observed that the time integral of the magnetic part of the momentum flux density $\mathcal{T}_B$ predicted by soliton theory matches well the mechanical momentum loss observed in simulations upon reflection, the basic reason for this agreement needs explanation. Indeed, while Eq.~(\ref{Eq:mom_conservation_all}) demonstrates that the time integral of the sum of the electric and magnetic parts of the momentum flux density is equal to the change in the sum of mechanical plus EM momentum, it does not in itself impose any particular partitioning. However, one shows from Maxwell's equation that
\begin{equation}
-\frac{\partial}{\partial t}\left[\epsilon_0E_yB_z\right] +\frac{\partial}{\partial x}T_{xx} = \rho_e E_x + j_y B_z. 
\label{Eq:momentum_conservation_lorentz}
\end{equation}
For the momentum partitioning Eq.~(\ref{Eq:momentum_partitioning}) observed in simulations to hold the volume integrated Lorentz force should then be equal to the magnetic part of the momentum flux density $\mathcal{T}_B$.

\subsubsection{Volume integrated Lorentz force in a soliton}

The local conservation equation for mechanical momentum along $\mathbf{\hat{x}}$ can be integrated over the simulation volume to yield
\begin{equation}
\label{Eq:mecha_momentum_conservation}
\frac{d}{dt} \mathpzc{Q}\,_{sp} = \int_V [\rho E_x + j_y B_z] dx.
\end{equation}
Recalling that the transverse current is primarily carried by the electron and that $v_{e_y}\sim -E_x/B_z$, one can assume $j_yB_z\sim e n_e E_x$. In addition, since $|n_i-n_e|\ll n_i$, the electric force $e(n_i-n_e)E_x$ is small compared to the Laplace force $j_yB_z$. The Lorentz force $\rho_e E_x + j_y B_z$ on the right hand side of Eq.~(\ref{Eq:mecha_momentum_conservation}) is thus approximately $e n_i E_x$. 

\begin{figure}
\begin{center}
\includegraphics[]{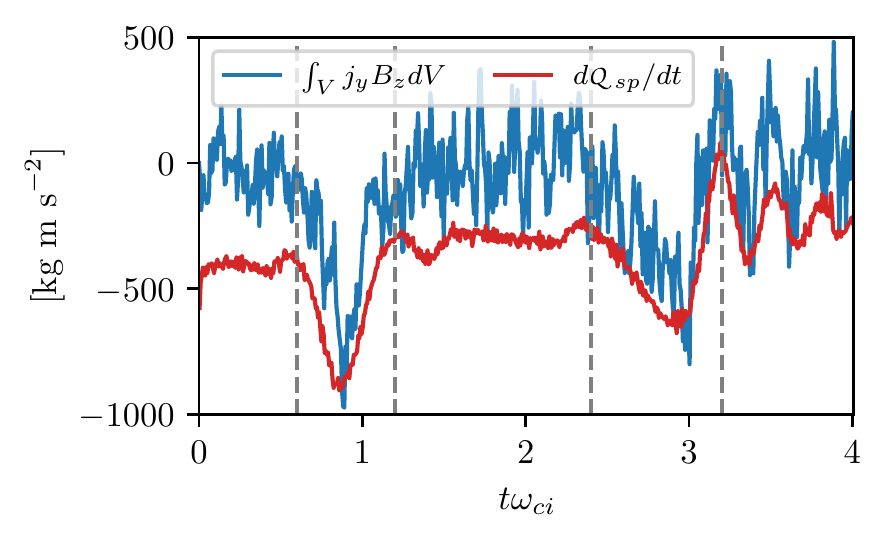}
\caption{Time evolution of rate of change of mechanical longitudinal momentum $\mathpzc{Q}\,_{sp}$ and volume integrated Laplace force $j_yB_z$. The Laplace force is observed to approach the rate of change of $\mathpzc{Q}\,_{sp}$ during the reflection events, that is for $0.6\leq t\omega_{ci}\leq1.2$ and $2.4\leq t\omega_{ci}\leq3.2$. }
\label{Fig:JyBz}
\end{center}
\end{figure}

For the soliton pulse profiles given in Eq.~(\ref{Eq:Soliton_scaling}), $E_x$ and $n_i$ are respectively odd and even function with respect to the pulse center in an infinite plasma. The Laplace force is hence odd and its volume integrated $\int_V j_yB_z dV$ is thus zero. This is consistent with the approximately zero volume integrated Laplace force observed in numerical simulations in Fig.~\ref{Fig:JyBz}.

This picture, however, is altered as the pulse approaches a plasma-vacuum interface in which case the volume integrated Lorentz force can momentarily become non-zero. This non-zero Lorentz force finds its origin, as we shall see, in the co-existence of a right propagating compression pulse and a left-propagating rarefaction pulse. 

\subsubsection{Longitudinal electric field during reflection}

\begin{figure}
\begin{center}
\includegraphics[]{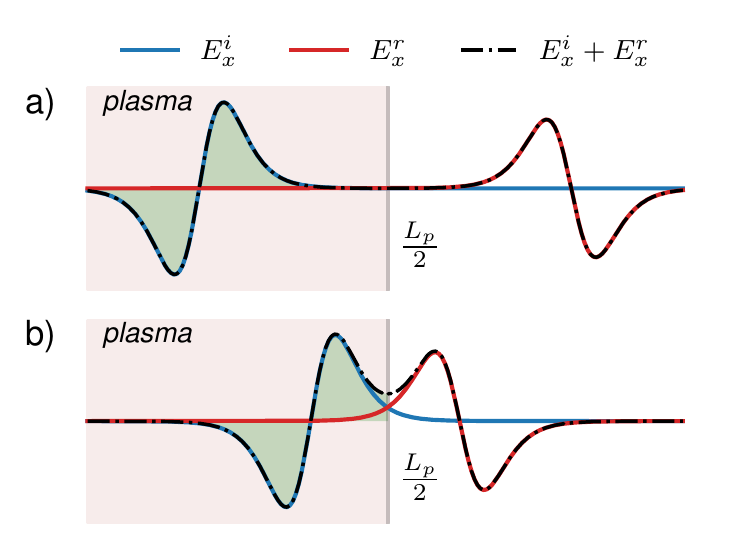}
\caption{Illustration of the interference of incident and reflected longitudinal electric fields. a) Away from reflection the volume integral of $E_x$ in the plasma slab $-L_p/2\leq x\leq L_p/2$ (olive area) is zero. b) During reflection the incident and reflected pulses interfere so that, for incident and reflected pulses of different amplitudes, this integral is no longer zero. }
\label{Fig:CombinedFields}
\end{center}
\end{figure}

As the pulse is reflected at the interface, both the longitudinal electric field and the density perturbation are distorted. As illustrated in Fig.~\ref{Fig:CombinedFields}, a simplified picture for the resulting field can be obtained by considering the superposition of an incident and a reflected pulse. Note here though that while we consider hypothetical longitudinal electric fields extending further than the plasma slab, this is only a mathematical model and the physical longitudinal field is as expected zero outside of the plasma slab. We consider here the reflection of a compression pulse at $x=L_p/2$, even if as we will show a similar argument can be derived for the reflection of a rarefaction pulse at $x=-L_p/2$. We also limit ourselves, for the sake of simplicity, to the effect of the combined incident and reflected longitudinal electric fields in a constant plasma density $n_0$. For completeness one should also consider incident and reflected perturbations in density, as well as cross terms, in the force density $en_i E_x$.

From Eq.~(\ref{Eq:Soliton_scaling}), the longitudinal field associated with a compression pulse of amplitude $\epsilon_i$ propagating towards $\mathbf{\hat{x}}$ writes 
\begin{equation}
E_x^i = \frac{\epsilon_i^{3/2}}{\eta}v_AB_0(\epsilon)\textrm{sech}^2\left[\psi_{\epsilon_i}^+(x,t)\right]\textrm{tanh}\left[\psi_{\epsilon_i}^+(x,t)\right],
\end{equation}
with $\psi_{\epsilon_i}^+(x,t) = \omega_{ce}\sqrt{\epsilon_i}(x-v_At)/(2c)$. Similarly, and having in mind reflection at $x=L_p/2$, we define 
\begin{equation}
E_x^r = -\frac{\epsilon_r^{3/2}}{\eta}v_AB_0\textrm{sech}^2\left[\psi_{\epsilon_r}^-(x,t)\right]\textrm{tanh}\left[\psi_{\epsilon_r}^-(x,t)\right],
\end{equation} 
with $\psi_{\epsilon_r}^-(x,t) = \omega_{ce}\sqrt{\epsilon_r}(x-L_p+v_At)/(2c)$ the longitudinal field associated with a rarefaction pulse of amplitude $\epsilon_r$ propagating towards $-\mathbf{\hat{x}}$. Over the course of the reflection, one must consider the force density $e n_i E_x$ where $E_x$ is the sum of incident and reflected fields. 

For a perfect reflection, that is $\epsilon_r = \epsilon_i$, one gets by symmetry that
\begin{equation}
\int_0^{L_p/V_a}\int_{-\infty}^{L_p/2} e n_0 (E_x^i+E_x^r) dxdt = 0.
\end{equation}
From Eq.~(\ref{Eq:mecha_momentum_conservation}) the result that the time and volume integral of the force density $en_0E_x$ is zero implies that there is no change in mechanical momentum. 

Consider now a partial reflection such that $\epsilon_r = r_F \epsilon_i$ with  $r_F\sim(1-2v_A/c)$ the reflection coefficient for the magnetic field. Remarking that
\begin{equation}
\int_0^{L_p/V_a}\int_{-\infty}^{L_p/2} e n_0 E_x^i dxdt = |\hat{\mathpzc{Q}\,}|,
\end{equation} 
one then gets to lowest order in $v_A/c$
\begin{equation}
\label{Eq:Ex_momentum_change}
\int_0^{L_p/V_a}\int_{-\infty}^{L_p/2} e n_0 (E_x^i+E_x^r) dxdt = -\frac{v_A}{c}|\hat{\mathpzc{Q}\,}|.
\end{equation}
In contrast with perfect reflection, there is now a change in mechanical momentum $\mathpzc{Q}\,_{sp}$. Quantitatively, the $v_A/c\ll1$ scaling obtained here is consistent with the magnitude of the loss in mechanical momentum observed in simulations upon reflection. The prediction here based on interfering fields that a mechanical momentum loss only occurs for a partial reflection is consistent with the results obtained above by Taylor expanding the momentum flux density at the simulation boundary, that is $\mathcal{T}_B^0=0$ but $\mathcal{T}_B^1\neq0$. Note finally that this simple picture of overlapping incident and reflected longitudinal electric fields also holds for the loss of mechanical momentum at the left boundary when transforming a rarefaction pulse into a compression pulse, as observed numerically. Indeed, in both cases, the integral of the reflected pulse in the plasma volume $\int_{-L_p/2}^{L_p/2}E_x^rdx$ does not fully compensate for the part of the incident pulse past the plasma edge $\int_{L_p/2}^{\infty}E_x^idx$ (or $\int_{-\infty}^{-L_p/2}E_x^idx$ at the left boundary) because of the smaller amplitude of the reflected pulse. This in turn leads to a negative force, consistent with the observed mechanical momentum loss. 

The momentum loss predicted in Eq.~(\ref{Eq:Ex_momentum_change}) does, however, differ by a factor $2$ from the result estimated in Fig.~\ref{Fig:Momentum} and derived from $\mathcal{T}_B^1$ in Eq.~(\ref{Eq:P1B}). The fact that Eq.~(\ref{Eq:Ex_momentum_change}) does not fully account for the measured loss in momentum should not be a surprise though since, as explained above, we limited our analysis here to the effect of the combined incident and reflected longitudinal electric fields. It stands to reason that accounting for the extra contributions due to density perturbations and cross terms will yield the change in momentum $\mathcal{P}^{1,B}$ predicted in Eq.~(\ref{Eq:P1B}). This hypothesis is further supported by the observation that, as shown in Fig.~\ref{Fig:JyBz}, the volume integrated Lorentz force does match the rate of change of $\mathpzc{Q}\,_{sp}$ during reflection events. 

To summarize, the analytical results obtained here for soliton-like pulse support that the time integral of the magnetic part of the momentum flux density is indeed associated with the volume and time integrated Lorentz force, confirming in turn the momentum flux partitioning observed in simulations.

\section{Summary}
\label{Sec:Summary}

In this paper we examine energy and momentum conservation when a magnetosonic (MS) pulse bounces back and forth in a finite width magnetized plasma slab bounded on both ends by vacuum. 

Analytical formulae for the mechanical and field contributions to the energy and longitudinal momentum in this systems are derived for a MS soliton-like pulse. For a small amplitude soliton energy is shown to be equally partioned between field and mechanical components, whereas compression and rarefaction configurations are found to correspond respectively to a high- and a low-energy state for the pulse-slab system. In the meantime, mechanical momentum in the system is shown to be $(c/v_A)^2$ larger than its electromagnetic counterpart, while the sum of these two contributions matches Minkowski's momentum $\mathcal{G}_M$.

Using particle-in-cell simulations, the dynamics of pulse bouncing and its effect on energy and momentum is investigated. Analysis of simulations results confirms that the transformation of a pulse from compression to rarefaction and vice-versa leads to a change in the energy of the system, and that this change in energy matches the time integrated Poynting flux during pulse reflection at the interface. Pulse reflection at the interface and bouncing in the slab is thus accompanied by alternating gains and losses in energy. For a partial reflection, it is further found that the losses in mechanical and EM pulse energy both follow separately Fresnel's transmission coefficient for the energy $1-{r_F}^2$.

Looking at momentum, particle-in-cell results confirm that pulse reflection at the interface leads to a reflection of the pulse EM momentum, but not of the mechanical momentum. The reflected rarefaction pulse thus propagates in the direction opposite to the mechanical momentum associated with this pulse. For perfect reflection, the theoretical analysis of the momentum flux density in the vacuum region derived for soliton pulses predict no loss of mechanical momentum but a change in the pulse-slab EM longitudinal momentum reflecting exactly the change in pulse EM longitudinal momentum. For partial reflection at the interface, however, this same analysis predicts, in addition to the change in EM momentum, a loss in momentum that happens to match the momentum loss constructed from the incident pulse mechanical momentum and Fresnel's transmission coefficient for the magnetic perturbation. Interestingly, this prediction is consistent with the loss in mechanical momentum measured in numerical simulations. Since the mechanical momentum is much larger than its EM counterpart, and that the total momentum for the system matches Minkowski's momentum, this suggests a loss of momentum equal to $(1-r_F)\mathcal{G}_M$ at each partial reflection at the plasma-vacuum interface. 

Analysis of simulation results further shows that the time evolution of the mechanical and EM momentum in the system are respectively associated with the magnetic and electric parts of the momentum flux densities. It is interesting to note that this result matches what is predicted for a homogeneous isotropic dielectric in the limit of a large relative permittivity, possibly suggesting a broader validity for this momentum partitioning.

Finally, simulations exposed a steady loss of mechanical momentum away from reflection events for which no satisfactory explanation has yet been found. It has been shown that this loss is remarkably consistent with the change in longitudinal momentum a single ion experience in the wake of the travelling pulse on timescales short compared to the ion gyro-period, but that this mechanism cannot explain the linear time dependence. The investigation of this question is left for future studies.

Looking ahead, an interesting extension of this work would be to examine the case of a step up in density at the discontinuity, so as to match the conditions for which Nagasawa~\cite{Nagasawa1986} predicted the possibility to reflect a soliton. Beyond its basic interest, this question is relevant for the problem of fast magnetic compression where the MS soliton propagating ahead of one shock front~\cite{Gueroult2017} could interact with the density step up associated with the counter-propagating shock~\cite{Gueroult2016}. Finally, since as hinted at by Ohsawa~\cite{Ohsawa2017} there exist strong similarities between the generation of MS solitons in fast magnetic compression experiments~\cite{Gueroult2017,Gueroult2016} and when a charged particles bunch penetrates in a plasma~\cite{Kumar2020}, this problem may also find applications in astrophysical settings.

\section*{Acknowledgments}

It is a pleasure for the author to thank Pr. Amnon Fruchtman for constructive discussions. 

This work was granted access to the HPC resources of CALMIP supercomputing center under the allocations 2019-p18015 and 2020-p18015.

\appendix

\section{Longitudinal ion velocity in soliton pulse}
\label{Sec:App1}

The momentum equation for an ion writes
\begin{subequations}
\label{Eq:ion_momentum}
\begin{align}
m_p\frac{dv_{i_x}}{dt} & = e \left[E_x(x_i(t),t) +v_{i_y} B_z(x_i(t),t)\right],\\
m_p\frac{dv_{i_y}}{dt} & = e \left[E_y(x_i(t),t) -v_{i_x} B_z(x_i(t),t)\right].
\end{align}
\end{subequations}
In an overdense plasma and for a small amplitude pulse we have previously shown that the ion longitudinal displacement resulting from the pulse passage is $\epsilon$ smaller than the pulse width~\cite{Gueroult2018b}. To lowest order the field quantities on the right-hand side of Eq.(\ref{Eq:ion_momentum}) can hence simply be evaluated at the ion initial position (that is $x_i(t)\sim x_i(t=0)$), in which case one obtains for the soliton pulse governed by Eq.~(\ref{Eq:Soliton_scaling})
\begin{subequations}
\label{Eq:ion_dynamics}
\begin{align}
\dot{\upsilon_x} & = 2[\textrm{sech}^2(-\bar{t})\textrm{tanh}(-\bar{t})+\frac{\eta}{\sqrt{\epsilon}}\upsilon_y[1+\epsilon\textrm{sech}^2(-\bar{t})]],\\
\dot{\upsilon_y} & = 2\frac{\eta}{\sqrt{\epsilon}}\left[\textrm{sech}^2(-\bar{t})-\upsilon_x[1+\epsilon\textrm{sech}^2(-\bar{t})]\right],
\end{align}
\end{subequations}
where time and speed have been normalized by $2\eta/(\omega_{ci}\sqrt{\epsilon})$ and $\epsilon v_A$, respectively, and $\dot{x}$ is the derivative of $x$ with respect to the dimensionless time $\bar{t}$.

To lowest order in $\epsilon$, the additional $\epsilon\textrm{sech}^2(-t)$ in the Lorentz force in the bracketed term on the right hand side which arises from the magnetic perturbation can be neglected. In this case, the two first order differential equation given in Eq.~(\ref{Eq:ion_dynamics}) can be combined in a single second order equation for $\upsilon_x$,
\begin{equation}
\ddot{\upsilon_x}+\frac{4\eta^2}{\epsilon}\upsilon_x = 2\frac{d}{d\bar{t}}[\textrm{sech}^2(-\bar{t})\textrm{tanh}(-\bar{t})]\\+\frac{4\eta^2}{\epsilon}\textrm{sech}^2(-\bar{t}).
\end{equation}
The ion velocity then writes 
\begin{subequations}
\label{Eq:low_order_sol}
\begin{align}
\upsilon_x(\bar{t}) & = \textrm{sech}^2(\bar{t})+\Upsilon\cos\left(\frac{2\eta}{\sqrt{\epsilon}}\bar{t}+\phi\right),\\
\upsilon_y(\bar{t}) & = -\Upsilon\sin\left(\frac{2\eta}{\sqrt{\epsilon}}\bar{t}+\phi\right),
\end{align}
\end{subequations}
with $\Upsilon$ and $\phi$ constants to be determined through initial conditions. To this order the ion motion is thus identical to that of an unmagnetized ion interacting with the longitudinal field only superimposed to the slower ion cyclotron motion. In other words the effects of the transverse field $E_y$ and the transverse Lorentz force $v_x B_0$ in the pulse cancel out.

\begin{figure}
\begin{center}
\includegraphics[]{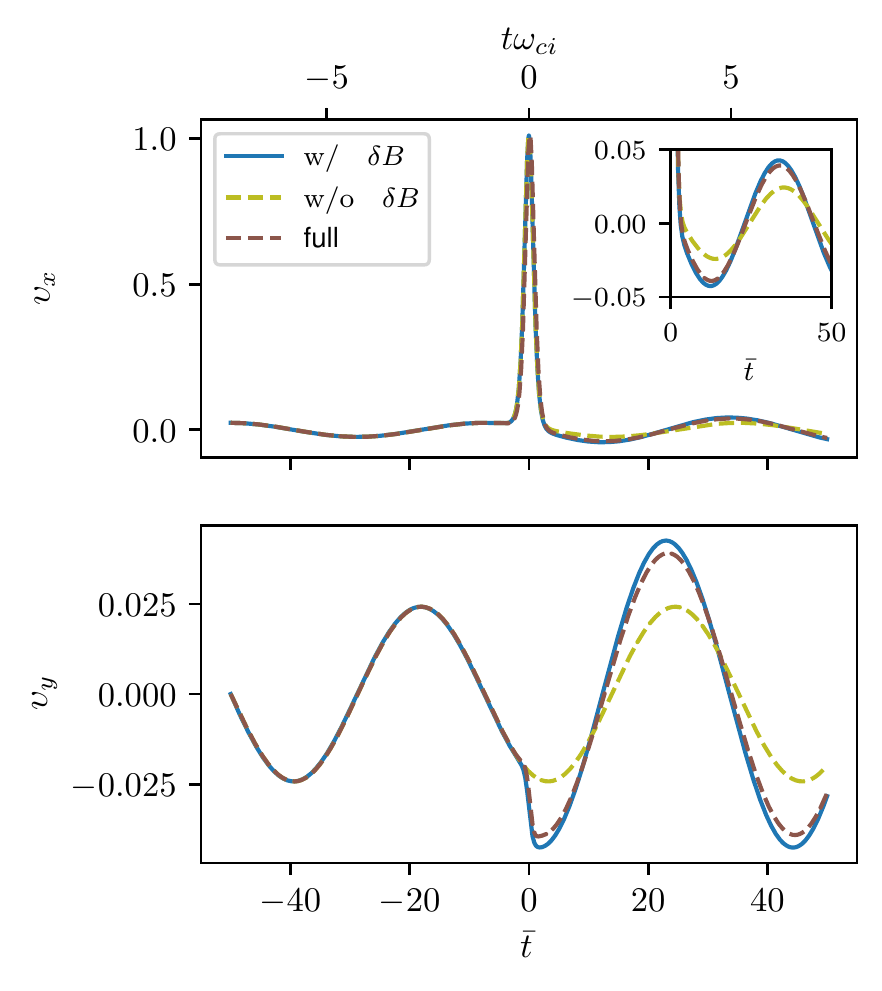}
\caption{Time evolution of the dimensionless longitudinal and transverse ion velocity $\upsilon_x = v_{i_x}/(\epsilon v_A)$ and $\upsilon_y = v_{i_y}/(\epsilon v_A)$ in response to the soliton electromagnetic field ($E_x,E_y,B_z$) with [Eq.~(\ref{Eq:full_solution})] and without [Eq.~(\ref{Eq:low_order_sol})] considering the effect of the magnetic perturbation $\delta B$, and solving numerically Eq~(\ref{Eq:ion_momentum}) using a Boris scheme (full). In all cases the ion is initialized with thermal velocity directed along $\mathbf{\hat{x}}$ at $\bar{t}=-50$, that is much before the soliton field is noticeable at the ion position. The soliton field peaks at the ion positon at t=0. Conditions are those of the PIC simulations given in Table~\ref{Tab:input_deck}. Upper and lower x-axes are related through $2\eta\bar{t}/\sqrt{\epsilon}=t\omega_{ci}$. }
\label{Fig:IonTrajSoliton}
\end{center}
\end{figure}

Looking now for a solution accounting for the magnetic perturbation, the system of coupled differential equations Eq.~(\ref{Eq:ion_dynamics}) has exact solutions in the form of
\begin{widetext}
\begin{subequations}
\label{Eq:full_solution}
\begin{align}
\upsilon_x(\bar{t}) & = \cos\left[\tau(\bar{t})\right]\left[\mathcal{C}_1-\frac{2\eta}{\sqrt{\epsilon}}\int_1^{\bar{t}}\textrm{sech}^2(u)\sin\left[\tau(u)\right]du-2\int_1^{\bar{t}}\textrm{sech}^2(u)\cos\left[\tau(u)\right]\textrm{tanh}(u)du\right]\nonumber\\
& + \sin\left[\tau(\bar{t})\right] \left[\mathcal{C}_2+\frac{2\eta}{\sqrt{\epsilon}}\int_1^{\bar{t}}\textrm{sech}^2(u)\cos\left[\tau(u)\right]du-2\int_1^{\bar{t}}\textrm{sech}^2(u)\sin\left[\tau(u)\right]\textrm{tanh}(u)du\right],\label{Eq:full_solution_x}\\
\upsilon_y(\bar{t}) & = -\sin\left[\tau(\bar{t})\right]\left[\mathcal{C}_1-\frac{2\eta}{\sqrt{\epsilon}}\int_1^{\bar{t}}\textrm{sech}^2(u)\sin\left[\tau(u)\right]du-2\int_1^{\bar{t}}\textrm{sech}^2(u)\cos\left[\tau(u)\right]\textrm{tanh}(u)du\right]\nonumber\\
& + \cos\left[\tau(\bar{t})\right] \left[\mathcal{C}_2+\frac{2\eta}{\sqrt{\epsilon}}\int_1^{\bar{t}}\textrm{sech}^2(u)\cos\left[\tau(u)\right]du-2\int_1^{\bar{t}}\textrm{sech}^2(u)\sin\left[\tau(u)\right]\textrm{tanh}(u)du\right],
\end{align}
\end{subequations}
\end{widetext}
where we have introduced the scaled time variable $\tau(\bar{t}) = 2\eta/\sqrt{\epsilon} [\bar{t}+ \epsilon\textrm{tanh}(\bar{t})]$ and $\mathcal{C}_1$ and $\mathcal{C}_2$ are constants to be determined from initial conditions. The effect of the magnetic perturbation on the orbit of a test ion is illustrated in Fig.~\ref{Fig:IonTrajSoliton}. We verify in passing that the solution in dotted-brown obtained by directly solving numerically Eq~(\ref{Eq:ion_momentum}) is qualitatively consistent with the result from Eq.~(\ref{Eq:full_solution}), supporting the assumption of negligible ion motion during the pulse passage.

\begin{figure}
\begin{center}
\includegraphics[]{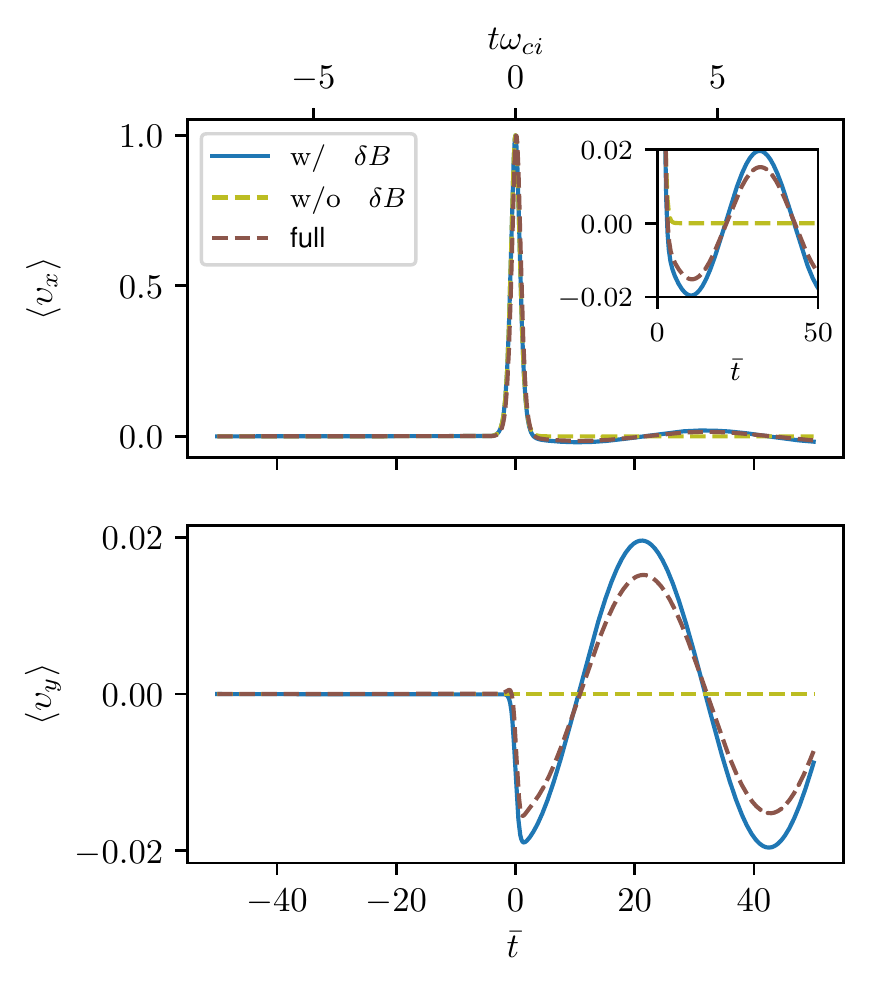}
\caption{Time evolution of the dimensionless longitudinal and transverse population averaged ion velocity $\langle\upsilon_x\rangle = \langle v_{i_x}\rangle/(\epsilon v_A)$ and $\langle\upsilon_y\rangle = \langle v_{i_y}\rangle(\epsilon v_A)$ in response to the soliton electromagnetic field ($E_x,E_y,B_z$) with [Eq.~(\ref{Eq:full_solution})] and without [Eq.~(\ref{Eq:low_order_sol})] considering the effect of the magnetic perturbation $\delta B$, and solving numerically Eq~(\ref{Eq:ion_momentum}) using a Boris scheme (full). In all cases we consider ions initialized with randomly directed thermal velocity at $\bar{t}=-50$, that is much before the soliton field is noticeable at the ion position. The soliton field peaks at the ion positon at $t=0$. Conditions are those of the PIC simulations given in Table~\ref{Tab:input_deck}. Upper and lower $x$-axes are related through $2\eta\bar{t}/\sqrt{\epsilon}=t\omega_{ci}$. }
\label{Fig:MeanIonTrajSoliton}
\end{center}
\end{figure}

Consider now a mono-energetic ion population ahead of the pulse with energy $m_p[\Upsilon\epsilon v_A]^2/2$. Since the gyro-phase of these ions is evenly distributed over $[0,2\pi]$, the population-averaged momentum $m_p\langle\upsilon_x\rangle$ is zero. On the other hand, Eq.~(\ref{Eq:full_solution_x}) indicates that the pulse passage introduces an average non-zero momentum $\langle\Delta p_x\rangle$ along $x$. This result is confirmed in Fig.~\ref{Fig:MeanIonTrajSoliton}. For $\bar{t}\gg1$, that is far downstream of the pulse, one finds
\begin{widetext}
\begin{equation}
\label{Eq:DeltaPx}
\langle\Delta p_x\rangle(\bar{t}) = 4 m_p\epsilon v_A\sin\left[\tau(\bar{t})\right]  \int_{0}^{\infty}\textrm{sech}^2(u)\left(\frac{\eta}{\sqrt{\epsilon}}\cos\left[\tau(u)\right]-\sin\left[\tau(u)\right]\textrm{tanh}(u)\right)du. 
\end{equation}
\end{widetext}
Clearly the time-average of $\langle\Delta p_x\rangle$ over an ion cyclotron period is zero for $\bar{t}\gg1$ but its instantaneous value is not. Taylor expanding the term in parentheses under the integral in Eq.~(\ref{Eq:DeltaPx}) for $\eta/\sqrt{\epsilon}\ll1$, the lowest order term is proportional to $1-u\textrm{tanh}(u)$ whose integral is zero. The next order term comes from the $\epsilon\textrm{tanh}(u)$ term in $\sin\left[\tau(u)\right]$ and is equal to $2\eta\sqrt{\epsilon}\textrm{tanh}^2(u)$. Since
\begin{equation}
\int \textrm{sech}^2(u)\textrm{tanh}^2(u)du = \frac{1}{3}\textrm{tanh}^3(u),
\end{equation}  
one gets 
\begin{equation}
\langle\Delta p_x\rangle\left(t\gg\frac{\sqrt{\epsilon}}{2\eta\omega_{ci}}\right)=-\frac{8}{3}\eta\epsilon^{3/2} m_p v_A \sin\left(\omega_{ci}t\right)
\end{equation}
to lowest order in $\eta/\sqrt{\epsilon}$. For $\epsilon=0.1$ and a physical mass ratio, $8\eta\sqrt{\epsilon}/3\sim2~10^{-2}$, which is consistent with the amplitude of $\langle\upsilon_x\rangle$ observed for $\bar{t} \gg1$ in Fig.~\ref{Fig:MeanIonTrajSoliton}. Note that the same result, or rather the same amplitude, can be obtained by simply considering the effect of the Lorentz force due to the magnetic perturbation $v_{i_x}B_1$ over the pulse passage. Indeed, integrating this force leads to a change in dimensionless transverse velocity 
\begin{equation}
\Delta \upsilon_y = -2\eta\sqrt{\epsilon}\int_{-\infty}^{\infty} \textrm{sech}^4(-\bar{t})d\bar{t} = -\frac{8}{3}\eta\sqrt{\epsilon}, 
\end{equation}
or in dimensional units $\Delta v_{i_y}=-8\eta\epsilon^{3/2}v_A/3$. This is consistent with the fact that now, as opposed to the situation in Eq.~(\ref{Eq:low_order_sol}), the larger Lorentz force exceeds the opposed electric force due to the transverse electric field. We also verify in Fig.~\ref{Fig:MeanIonTrajSoliton} that this behaviour is qualitatively recovered when solving directly Eq.~(\ref{Eq:ion_momentum}).

Since the pulse travels with a velocity $v_A$, the number of ions passing through the pulse per unit of time is $v_A n_0$. The decrease in momentum integrated over the ion population is hence 
\begin{equation}
\Delta \hat{\mathpzc{Q}\,}_p (t) = \int_0^t n_0v_A\Delta p_x(t')dt'.
\end{equation}
From Eq.~(\ref{Eq:momentum_Mech}), the normalized rate of change for the total longitudinal momentum in the system writes
\begin{equation}
\frac{\hat{\mathpzc{Q}\,}_p-\Delta \hat{\mathpzc{Q}\,}_p (t)}{\hat{\mathpzc{Q}\,}_p} = 1-\frac{2}{3}\epsilon\left[1-\cos(\omega_{ci}t)\right].
\end{equation}
We note, remarkably, that the order of magnitude of this effect is consistent with the decrease in momentum observed in Fig.~\ref{Fig:LossScaling}. Yet, the cosine time evolution obtained here does not match the linear decrease observed in simulations. 

Although this difference can not be fully accounted for, one can bring forward elements that could contribute to it. First, this calculation assumes that the ion population at $t=0$ is entirely upstream of the pulse, which is inconsistent with particle-in-cell simulation results. Indeed in PIC simulations the pulse is initially in the plasma slab and ions are initialized by adding a thermal velocity to the zero$^{th}$ order pulse longitudinal velocity. Ions that are already in the pulse at $t = 0$ will hence only experience a fraction of the effect modelled here. This is anticipated to be all the more important since the plasma slab width, that is the distance travelled by the pulse, is only a few soliton widths large. 

We must also note that this explanation is inconsistent with the decrease of longitudinal momentum observed for the reflected pulse. Indeed, because in this case the sign of the Lorentz force due to the magnetic perturbation is reversed as $B_1<0$, one finds $\Delta\upsilon_y>0$ and, as a result, $\langle\Delta p_x\rangle>0$. The inability of this model to capture the momentum evolution for the rarefaction pulse should however not be a surprise. This model indeed relies directly on EM fields matching soliton profiles, whereas it has been clearly shown in simulations and predicted theoretically that the reflected rarefaction pulse is not a soliton.

\section{Momentum partitioning from constitutive relations in a dielectric in the absence of free charges}
\label{Sec:App5}

Conservation of momentum in a medium can be written as~\cite{Jackson1970,Frias2012} 
\begin{equation}
\frac{\partial}{\partial t} \left[\epsilon_0 \mathbf{E}\times\mathbf{B}\right]  + \frac{d\mathbf{p}_m}{dt} = \bm{\nabla}\cdot \left(\mathbf{T^{E}}+\mathbf{T^{B}}\right),
 \label{Eq:vacuum}
\end{equation}
with
\begin{subequations}
\begin{equation}
T_{ij}^{E}  = \epsilon_0\left[E_iE_j-\frac{1}{2}E^2\delta_{ij}\right]
\end{equation}
and
\begin{equation}
T_{ij}^{B} = {\mu_0}^{-1}\left[B_iB_j-\frac{1}{2}B^2\delta_{ij}\right]
\end{equation}
\end{subequations}
the electric and magnetic part of the usual Maxwell stress tensor, and $\mathbf{p}_m$ the mechanical momentum density such that 
\begin{equation}
\frac{d \mathbf{p}_m}{dt} = \rho \mathbf{E} + \bm{j}\times\mathbf{B}.
\end{equation}

In a linear non-absorptive, non-dispersive and non-magnetic dielectric medium in the absence of free charges and currents, the source terms write
\begin{subequations}
\begin{equation}
\rho = -\bm{\nabla}\cdot (\epsilon_0\bm{\chi}\cdot \mathbf{E})
\end{equation}
and
\begin{equation}
\bm{j} =   \frac{\partial}{\partial t} \left[\epsilon_0\bm{\chi}\cdot \mathbf{E}\right],
\end{equation}
\end{subequations}
with $\bm{\chi}$ the susceptibility tensor such that 
\begin{equation}
\mathbf{D} = \epsilon_0(\mathds{1}+\bm{\chi})\mathbf{E}.
\end{equation}
The time derivative of the mechanical momentum density can then be rewritten in terms of electromagnetic field quantities only to yield
\begin{align}
\frac{d \mathbf{p}_m}{dt} & = -\left[\bm{\nabla}\cdot (\epsilon_0\bm{\chi}\cdot \mathbf{E})\right] \mathbf{E} + \frac{\partial}{\partial t} \left[\epsilon_0\bm{\chi}\cdot \mathbf{E}\right]\times\mathbf{B}\nonumber\\
 & = \left[\bm{\nabla}\cdot (\epsilon_0\bm{\chi}\cdot \mathbf{E})\right] \mathbf{E} + \frac{\partial}{\partial t} \left(\left[\epsilon_0\bm{\chi}\cdot \mathbf{E}\right]\times\mathbf{B}\right)\nonumber\\
 & \quad + \left[\epsilon_0\bm{\chi}\cdot \mathbf{E}\right]\times\bm{\nabla}\times\mathbf{E},
\end{align}
where we have used $\partial \mathbf{B}/\partial t = -\bm{\nabla}\mathbf{E}$ to obtain the last result. 

In the limit of a scalar uniform susceptibility $\bm{\chi} = \chi\mathds{1}$ and recalling that 
\begin{equation}
\frac{1}{2}\bm{\nabla}\left(\mathbf{A}\cdot\mathbf{A}\right) = \mathbf{A}\times(\bm{\nabla}\times\mathbf{A})+\left(\mathbf{A}\cdot\bm{\nabla}\right)\mathbf{A},
\end{equation}
one obtains from Eq.~(\ref{Eq:vacuum})
\begin{align}
\frac{d \mathbf{p}_m}{dt} & = \chi\left[\frac{\partial}{\partial t} \left(\epsilon_0 \mathbf{E}\times\mathbf{B}\right)-\bm{\nabla}\cdot \mathbf{T^{E}} \right]\nonumber\\
 & =  \frac{\epsilon_R-1}{\epsilon_R}\bm{\nabla}\cdot \mathbf{T^{B}}
\label{Eq:dielectric_momentum}
\end{align}
with $\epsilon_R = 1+\chi$ the relative permittivity. This last result is consistent with that obtained in a similar manner by Bisognano~\cite{Bisognano2017}. For $\epsilon_R\gg1$ Eq.~(\ref{Eq:dielectric_momentum}) shows that the variation of the mechanical momentum density is governed by the flux of the magnetic part of the Maxwell stress tensor.

\section*{References}
%

\end{document}